%% file: BELLE2-CONF-PH-2020-005.tex
\documentclass[preprint,prd,tightenlines,superscriptaddress]{revtex4-1}

\usepackage{graphicx} % Include figure files
\usepackage{dcolumn}  % Align table columns on decimal point
\usepackage{colordvi}
\usepackage{color}
\usepackage{epstopdf}
\usepackage{amssymb}
\usepackage{amsmath}
\usepackage{url}
\graphicspath{{ps}}
\usepackage{hyperref}
\usepackage{tabularx}
\usepackage{lineno}

\newcommand{\ifb}{\ensuremath{{\rm fb}^{-1}}\xspace}
\newcommand{\lumi}{\ensuremath{34.6~\ifb}\xspace}

\usepackage{pgf}
\usepackage{pgffor}
\usepackage{pgfplots}
\usepackage{tikz}
\usepackage{units}
\usepackage[super]{nth}
\usepackage{hepnames}

\usepackage{amssymb}
\usepackage{amsfonts}
\usepackage{upgreek} % Adds in support for greek letters in roman typeset
\usepackage[tikz]{bclogo}
\usetikzlibrary{pgfplots.colorbrewer,pgfplots.fillbetween}
\usetikzlibrary{shadows.blur}
\usetikzlibrary{shapes.symbols}
\usetikzlibrary{calc,intersections,matrix}
\usetikzlibrary{arrows,fadings,automata,snakes,shapes,shapes.misc,shapes.arrows,trees,positioning,calc,decorations.pathreplacing,arrows.meta}
\tikzfading[name=arrowfading, top color=transparent!0, bottom color=transparent!95]
\tikzset{arrowfill/.style={top color=blue!50, bottom color=blue,}}
\tikzset{>={Latex[width=1.5mm,length=1.5mm]}}
\tikzset{arrowstyle/.style={draw=black,arrowfill, single arrow,minimum height=#1, single arrow,
single arrow head extend=.4cm,}}

\tikzset{
    box/.style={
      rectangle,
	  color=#1,
      draw=black,
      fill=#1,
      thick,
      text=black,
      align=center,
      rounded corners=6pt,
      minimum height=1.5em
    }, 
    hbox/.style={
      rectangle,
      draw=black,
      fill=black,
      thick,
      text=white,
      align=center,
      rounded corners=6pt,
      minimum height=1.5em
    }, 
}
\colorlet{tree0}{blue}
\colorlet{tree100}{white!20!blue}
\colorlet{tree200}{white!40!blue}
\colorlet{tree300}{white!60!blue}
\colorlet{tree400}{white!80!blue}
\colorlet{tree500}{white}
\colorlet{tree600}{white!80!orange}
\colorlet{tree700}{white!60!orange}
\colorlet{tree800}{white!40!orange}
\colorlet{tree900}{white!20!orange}
\colorlet{tree1000}{orange}
\definecolor{Tblue}{HTML}{3465A4}	% tango sky blue 2
\definecolor{Tbluedark}{HTML}{204A87}	% tango sky blue 3
\definecolor{Tbluelight}{HTML}{729FCF}	% tango sky blue 1
\definecolor{Tbluelighter}{HTML}{8CC4FF}	% own definition of lighter blue
\definecolor{Tbrown}{HTML}{C17D11}	% tango chocolate 2
\definecolor{Tbrowndark}{HTML}{8F5902}	% tango chocolate 3
\definecolor{Tbrownlight}{HTML}{E9B96E}	% tango chocolate 1
\definecolor{Tgray}{HTML}{888A85}	% tango aluminium 4
\definecolor{Tgraydark}{HTML}{555753}	% tango aluminium 5
\definecolor{Tgraydarker}{HTML}{2E3436}	% tango aluminium 5
\definecolor{Tgraylight}{HTML}{BABDB6}	% tango aluminium 3
\definecolor{Tgraylight2}{HTML}{E4E6E2}	% Sehr hell (für Tabellenköpfe)
\definecolor{Tgraylight3}{HTML}{F0F2EE}	% Sehr hell (für Quelltexte)
\definecolor{Tgreen}{HTML}{73D216}	% tango chameleon 2	
\definecolor{Tgreendark}{HTML}{4E9A06}	% tango chameleon 3
\definecolor{Tgreenlight}{HTML}{8AE234}	% tango chameleon 1
\definecolor{Tred}{HTML}{CC0000}	% tango scarlet red 2
\definecolor{Treddark}{HTML}{A40000}	% tango scarlet red 3
\definecolor{Tredlight}{HTML}{EF2929}	% tango scarlet red 1	
\definecolor{Tlilac}{HTML}{75507B}	% tango plum 2
\definecolor{Tlilacdark}{HTML}{5C3566}	% tango plum 3
\definecolor{Tlilaclight}{HTML}{AD7FA8}	% tango plum 1
\definecolor{Tyellow}{HTML}{EDD400}	% tango butter 2
\definecolor{Tyellowdark}{HTML}{C4A000}	% tango butter 1
\definecolor{Tyellowlight}{HTML}{FCE94F}% tango butter 3
\definecolor{Torange}{HTML}{F57900}	% tango orange 2
\definecolor{Torangedark}{HTML}{CE5C00}	% tango orange 1
\definecolor{Torangelight}{HTML}{FCAF3E}% tango orange 3

\input ./belle2sym.tex

\begin{document}

%place for definitions and newcommands
\def\belletwo {\it {Belle II}}

\vspace*{-3\baselineskip}
\resizebox{!}{3cm}{\includegraphics{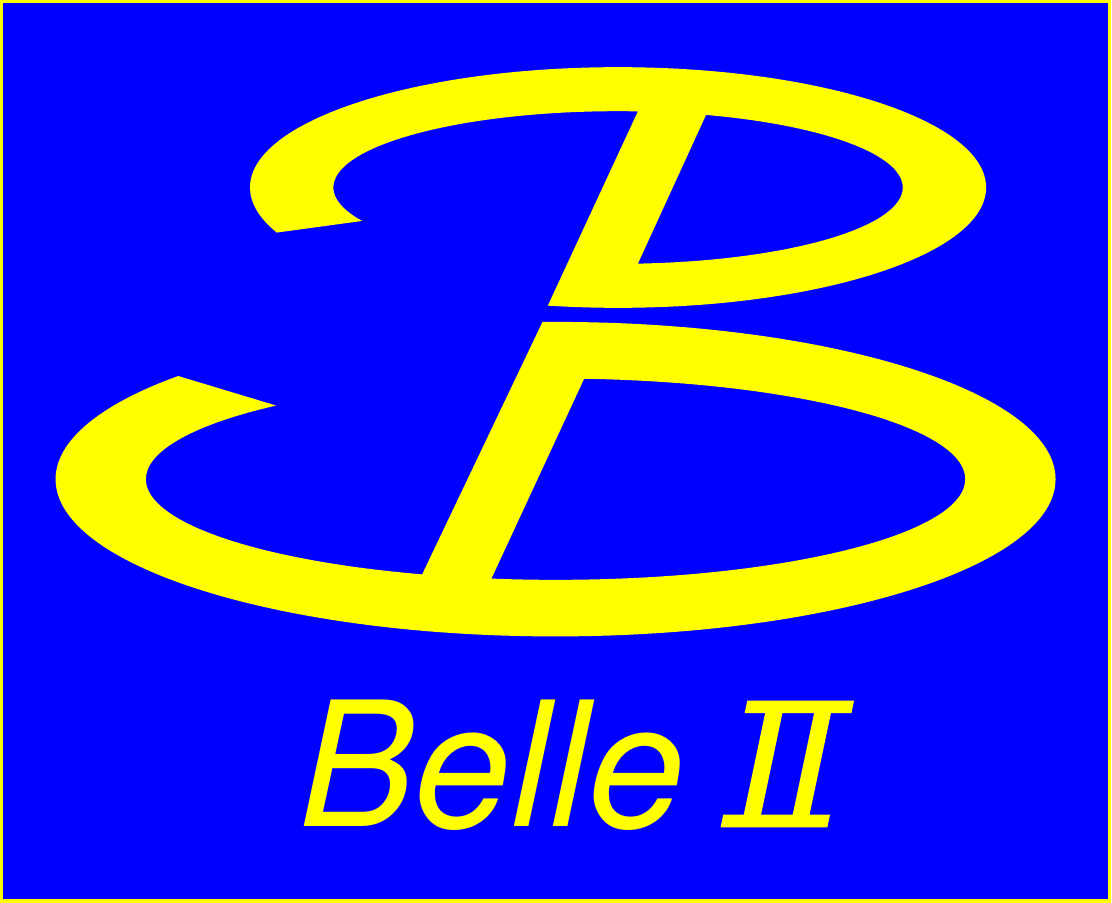}}

\vspace*{-5\baselineskip}
\begin{flushright}
BELLE2-CONF-PH-2020-005\\
\today
\end{flushright}

\title { \quad\\[0.5cm] A calibration of the Belle II hadronic tag-side reconstruction algorithm with $B \rightarrow X\ell \nu$ decays}

\input{authors-conf2020}

\collaboration{The Belle II Collaboration}
\noaffiliation

\begin{abstract}
	Tag-side reconstruction is an important method for reconstructing $B$ meson decays with missing energy. The Belle II tag-side reconstruction algorithm, Full Event Interpretation, relies on a hierarchical reconstruction of $B$ meson decays with multivariate classification employed at each stage of reconstruction. Given the large numbers of classifiers employed and decay chains reconstructed, the performance of the algorithm on data and simulation differs significantly. Here, calibration factors are derived for hadronic tag-side $B$ decays by measuring a signal side decay, $B \rightarrow X\ell \nu$, in $34.6$ fb$^{-1}$ of Belle II data. For a very loose selection on the tag-side $B$ multivariate classifier, the calibration factors are $0.65 \pm 0.02$ and $0.83 \pm 0.03$ for tag-side $B^{+}$ and $B^{0}$ mesons, respectively. 

\end{abstract}

\maketitle

{\renewcommand{\thefootnote}{\fnsymbol{footnote}}}
\setcounter{footnote}{0}

% Inputs here
\input{introduction}            % Owner: William
\input{detector_and_simulation} % Owner: William
\input{algorithm}               % Owner: William
\input{selection}             % Owner: William
\input{calibration}             % Owner: William
\input{systematics}             % Owner: William
\input{results}             % Owner: William
\input{conclusions}             % Owner: William

\section{ACKNOWLEDGEMENTS}
\input{acknowledgements}

\bibliography{belle2}
\bibliographystyle{belle2-note}
\input{appendix}             % Owner: William

\end{document}

%% file: authors-conf2020.tex
%%% Paper:    (2020  conference papers)
%%% Journal:  (2020 conferences)
%%% ====================================================================
%%% Use \input{authors-conf2020} to insert this material into your latex file.
\newcommand{\instSinica}{Academia Sinica, Taipei 11529, Taiwan}
\newcommand{\instCPPM}{Aix Marseille Universit\'{e}, CNRS/IN2P3, CPPM, 13288 Marseille, France}
\newcommand{\instBeihang}{Beihang University, Beijing 100191, China}
\newcommand{\instBUAP}{Benemerita Universidad Autonoma de Puebla, Puebla 72570, Mexico}
\newcommand{\instBNL}{Brookhaven National Laboratory, Upton, New York 11973, U.S.A.}
\newcommand{\instBINP}{Budker Institute of Nuclear Physics SB RAS, Novosibirsk 630090, Russian Federation}
\newcommand{\instCMU}{Carnegie Mellon University, Pittsburgh, Pennsylvania 15213, U.S.A.}
\newcommand{\instCinvestavIPN}{Centro de Investigacion y de Estudios Avanzados del Instituto Politecnico Nacional, Mexico City 07360, Mexico}
\newcommand{\instPrague}{Faculty of Mathematics and Physics, Charles University, 121 16 Prague, Czech Republic}
\newcommand{\instChiangMai}{Chiang Mai University, Chiang Mai 50202, Thailand}
\newcommand{\instChiba}{Chiba University, Chiba 263-8522, Japan}
\newcommand{\instChonnam}{Chonnam National University, Gwangju 61186, South Korea}
\newcommand{\instConacyt}{Consejo Nacional de Ciencia y Tecnolog\'{\i}a, Mexico City 03940, Mexico}
\newcommand{\instDESY}{Deutsches Elektronen--Synchrotron, 22607 Hamburg, Germany}
\newcommand{\instDuke}{Duke University, Durham, North Carolina 27708, U.S.A.}
\newcommand{\instITAR}{Institute of Theoretical and Applied Research (ITAR), Duy Tan University, Hanoi 100000, Vietnam}
\newcommand{\instENEA}{ENEA Casaccia, I-00123 Roma, Italy}
\newcommand{\instEri}{Earthquake Research Institute, University of Tokyo, Tokyo 113-0032, Japan}
\newcommand{\instJuelich}{Forschungszentrum J\"{u}lich, 52425 J\"{u}lich, Germany}
\newcommand{\instFuJen}{Department of Physics, Fu Jen Catholic University, Taipei 24205, Taiwan}
\newcommand{\instFudan}{Key Laboratory of Nuclear Physics and Ion-beam Application (MOE) and Institute of Modern Physics, Fudan University, Shanghai 200443, China}
\newcommand{\instGoettingen}{II. Physikalisches Institut, Georg-August-Universit\"{a}t G\"{o}ttingen, 37073 G\"{o}ttingen, Germany}
\newcommand{\instGifu}{Gifu University, Gifu 501-1193, Japan}
\newcommand{\instSOKENDAI}{The Graduate University for Advanced Studies (SOKENDAI), Hayama 240-0193, Japan}
\newcommand{\instGyeongsang}{Gyeongsang National University, Jinju 52828, South Korea}
\newcommand{\instHanyang}{Department of Physics and Institute of Natural Sciences, Hanyang University, Seoul 04763, South Korea}
\newcommand{\instKEK}{High Energy Accelerator Research Organization (KEK), Tsukuba 305-0801, Japan}
\newcommand{\instJPARC}{J-PARC Branch, KEK Theory Center, High Energy Accelerator Research Organization (KEK), Tsukuba 305-0801, Japan}
\newcommand{\instHSE}{Higher School of Economics (HSE), Moscow 101000, Russian Federation}
\newcommand{\instIISER}{Indian Institute of Science Education and Research Mohali, SAS Nagar, 140306, India}
\newcommand{\instIITBhubaneswar}{Indian Institute of Technology Bhubaneswar, Satya Nagar 751007, India}
\newcommand{\instIITGuwahati}{Indian Institute of Technology Guwahati, Assam 781039, India}
\newcommand{\instIITHyderabad}{Indian Institute of Technology Hyderabad, Telangana 502285, India}
\newcommand{\instIITMadras}{Indian Institute of Technology Madras, Chennai 600036, India}
\newcommand{\instIndiana}{Indiana University, Bloomington, Indiana 47408, U.S.A.}
\newcommand{\instIHEPRussia}{Institute for High Energy Physics, Protvino 142281, Russian Federation}
\newcommand{\instHEPHYVienna}{Institute of High Energy Physics, Vienna 1050, Austria}
\newcommand{\instIHEPChina}{Institute of High Energy Physics, Chinese Academy of Sciences, Beijing 100049, China}
\newcommand{\instChennai}{Institute of Mathematical Sciences, Chennai 600113, India}
\newcommand{\instIPP}{Institute of Particle Physics (Canada), Victoria, British Columbia V8W 2Y2, Canada}
\newcommand{\instIOP}{Institute of Physics, Vietnam Academy of Science and Technology (VAST), Hanoi, Vietnam}
\newcommand{\instIFIC}{Instituto de Fisica Corpuscular, Paterna 46980, Spain}
\newcommand{\instFrascati}{INFN Laboratori Nazionali di Frascati, I-00044 Frascati, Italy}
\newcommand{\instNapoliINFN}{INFN Sezione di Napoli, I-80126 Napoli, Italy}
\newcommand{\instPadovaINFN}{INFN Sezione di Padova, I-35131 Padova, Italy}
\newcommand{\instPerugiaINFN}{INFN Sezione di Perugia, I-06123 Perugia, Italy}
\newcommand{\instPisaINFN}{INFN Sezione di Pisa, I-56127 Pisa, Italy}
\newcommand{\instRomaINFN}{INFN Sezione di Roma, I-00185 Roma, Italy}
\newcommand{\instRomaTreINFN}{INFN Sezione di Roma Tre, I-00146 Roma, Italy}
\newcommand{\instTorinoINFN}{INFN Sezione di Torino, I-10125 Torino, Italy}
\newcommand{\instTriesteINFN}{INFN Sezione di Trieste, I-34127 Trieste, Italy}
\newcommand{\instJAEA}{Advanced Science Research Center, Japan Atomic Energy Agency, Naka 319-1195, Japan}
\newcommand{\instMainz}{Johannes Gutenberg-Universit\"{a}t Mainz, Institut f\"{u}r Kernphysik, D-55099 Mainz, Germany}
\newcommand{\instGiessen}{Justus-Liebig-Universit\"{a}t Gie\ss{}en, 35392 Gie\ss{}en, Germany}
\newcommand{\instKarlsruhe}{Institut f\"{u}r Experimentelle Teilchenphysik, Karlsruher Institut f\"{u}r Technologie, 76131 Karlsruhe, Germany}
\newcommand{\instKennesaw}{Kennesaw State University, Kennesaw, Georgia 30144, U.S.A.}
\newcommand{\instKitasato}{Kitasato University, Sagamihara 252-0373, Japan}
\newcommand{\instKISTI}{Korea Institute of Science and Technology Information, Daejeon 34141, South Korea}
\newcommand{\instKorea}{Korea University, Seoul 02841, South Korea}
\newcommand{\instKSU}{Kyoto Sangyo University, Kyoto 603-8555, Japan}
\newcommand{\instKyotoU}{Kyoto University, Kyoto 606-8501, Japan}
\newcommand{\instKyungpook}{Kyungpook National University, Daegu 41566, South Korea}
\newcommand{\instLPI}{P.N. Lebedev Physical Institute of the Russian Academy of Sciences, Moscow 119991, Russian Federation}
\newcommand{\instLNNU}{Liaoning Normal University, Dalian 116029, China}
\newcommand{\instLMU}{Ludwig Maximilians University, 80539 Munich, Germany}
\newcommand{\instLuther}{Luther College, Decorah, Iowa 52101, U.S.A.}
\newcommand{\instMNITJaipur}{Malaviya National Institute of Technology Jaipur, Jaipur 302017, India}
\newcommand{\instMPP}{Max-Planck-Institut f\"{u}r Physik, 80805 M\"{u}nchen, Germany}
\newcommand{\instMPGHLL}{Semiconductor Laboratory of the Max Planck Society, 81739 M\"{u}nchen, Germany}
\newcommand{\instMcGill}{McGill University, Montr\'{e}al, Qu\'{e}bec, H3A 2T8, Canada}
\newcommand{\instMETU}{Middle East Technical University, 06531 Ankara, Turkey}
\newcommand{\instMEPhI}{Moscow Physical Engineering Institute, Moscow 115409, Russian Federation}
\newcommand{\instNagoya}{Graduate School of Science, Nagoya University, Nagoya 464-8602, Japan}
\newcommand{\instNagoyaKMI}{Kobayashi-Maskawa Institute, Nagoya University, Nagoya 464-8602, Japan}
\newcommand{\instNagoyaIAR}{Institute for Advanced Research, Nagoya University, Nagoya 464-8602, Japan}
\newcommand{\instNaraWu}{Nara Women's University, Nara 630-8506, Japan}
\newcommand{\instUNAM}{National Autonomous University of Mexico, Mexico City, Mexico}
\newcommand{\instNTUTaiwan}{Department of Physics, National Taiwan University, Taipei 10617, Taiwan}
\newcommand{\instNUUTaiwan}{National United University, Miao Li 36003, Taiwan}
\newcommand{\instKrakow}{H. Niewodniczanski Institute of Nuclear Physics, Krakow 31-342, Poland}
\newcommand{\instNiigata}{Niigata University, Niigata 950-2181, Japan}
\newcommand{\instNSU}{Novosibirsk State University, Novosibirsk 630090, Russian Federation}
\newcommand{\instOkinawa}{Okinawa Institute of Science and Technology, Okinawa 904-0495, Japan}
\newcommand{\instOsakaCity}{Osaka City University, Osaka 558-8585, Japan}
\newcommand{\instRCNP}{Research Center for Nuclear Physics, Osaka University, Osaka 567-0047, Japan}
\newcommand{\instPNNL}{Pacific Northwest National Laboratory, Richland, Washington 99352, U.S.A.}
\newcommand{\instPanjab}{Panjab University, Chandigarh 160014, India}
\newcommand{\instPeking}{Peking University, Beijing 100871, China}
\newcommand{\instPanjabPAU}{Punjab Agricultural University, Ludhiana 141004, India}
\newcommand{\instRIKENMSL}{Meson Science Laboratory, Cluster for Pioneering Research, RIKEN, Saitama 351-0198, Japan}
\newcommand{\instRIKEN}{Theoretical Research Division, Nishina Center, RIKEN, Saitama 351-0198, Japan}
\newcommand{\instXavier}{St. Francis Xavier University, Antigonish, Nova Scotia, B2G 2W5, Canada}
\newcommand{\instSeoul}{Seoul National University, Seoul 08826, South Korea}
\newcommand{\instShandong}{Shandong University, Jinan 250100, China}
\newcommand{\instSPU}{Showa Pharmaceutical University, Tokyo 194-8543, Japan}
\newcommand{\instSoochow}{Soochow University, Suzhou 215006, China}
\newcommand{\instSoongsil}{Soongsil University, Seoul 06978, South Korea}
\newcommand{\instLjubljanaJSI}{J. Stefan Institute, 1000 Ljubljana, Slovenia}
\newcommand{\instKyiv}{Taras Shevchenko National Univ. of Kiev, Kiev, Ukraine}
\newcommand{\instTata}{Tata Institute of Fundamental Research, Mumbai 400005, India}
\newcommand{\instTUM}{Department of Physics, Technische Universit\"{a}t M\"{u}nchen, 85748 Garching, Germany}
\newcommand{\instECUTUM}{Excellence Cluster Universe, Technische Universit\"{a}t M\"{u}nchen, 85748 Garching, Germany}
\newcommand{\instTelAviv}{Tel Aviv University, School of Physics and Astronomy, Tel Aviv, 69978, Israel}
\newcommand{\instToho}{Toho University, Funabashi 274-8510, Japan}
\newcommand{\instTohoku}{Department of Physics, Tohoku University, Sendai 980-8578, Japan}
\newcommand{\instTitech}{Tokyo Institute of Technology, Tokyo 152-8550, Japan}
\newcommand{\instTokyoMetropolitan}{Tokyo Metropolitan University, Tokyo 192-0397, Japan}
\newcommand{\instUAS}{Universidad Autonoma de Sinaloa, Sinaloa 80000, Mexico}
\newcommand{\instNapoliUNIV}{Dipartimento di Scienze Fisiche, Universit\`{a} di Napoli Federico II, I-80126 Napoli, Italy}
\newcommand{\instNapoliUNIVA}{Dipartimento di Agraria, Universit\`{a} di Napoli Federico II, I-80055 Portici (NA), Italy}
\newcommand{\instPadovaUNIV}{Dipartimento di Fisica e Astronomia, Universit\`{a} di Padova, I-35131 Padova, Italy}
\newcommand{\instPerugiaUNIV}{Dipartimento di Fisica, Universit\`{a} di Perugia, I-06123 Perugia, Italy}
\newcommand{\instPisaUNIV}{Dipartimento di Fisica, Universit\`{a} di Pisa, I-56127 Pisa, Italy}
\newcommand{\instRomaUNIV}{Universit\`{a} di Roma ``La Sapienza,'' I-00185 Roma, Italy}
\newcommand{\instRomaTreUNIV}{Dipartimento di Matematica e Fisica, Universit\`{a} di Roma Tre, I-00146 Roma, Italy}
\newcommand{\instTorinoUNIV}{Dipartimento di Fisica, Universit\`{a} di Torino, I-10125 Torino, Italy}
\newcommand{\instTriesteUNIV}{Dipartimento di Fisica, Universit\`{a} di Trieste, I-34127 Trieste, Italy}
\newcommand{\instMontreal}{Universit\'{e} de Montr\'{e}al, Physique des Particules, Montr\'{e}al, Qu\'{e}bec, H3C 3J7, Canada}
\newcommand{\instIJCLab}{Universit\'{e} Paris-Saclay, CNRS/IN2P3, IJCLab, 91405 Orsay, France}
\newcommand{\instIPHC}{Universit\'{e} de Strasbourg, CNRS, IPHC, UMR 7178, 67037 Strasbourg, France}
\newcommand{\instAdelaide}{Department of Physics, University of Adelaide, Adelaide, South Australia 5005, Australia}
\newcommand{\instBonn}{University of Bonn, 53115 Bonn, Germany}
\newcommand{\instUBC}{University of British Columbia, Vancouver, British Columbia, V6T 1Z1, Canada}
\newcommand{\instCincinnati}{University of Cincinnati, Cincinnati, Ohio 45221, U.S.A.}
\newcommand{\instFlorida}{University of Florida, Gainesville, Florida 32611, U.S.A.}
\newcommand{\instHamburg}{University of Hamburg, 20148 Hamburg, Germany}
\newcommand{\instHawaii}{University of Hawaii, Honolulu, Hawaii 96822, U.S.A.}
\newcommand{\instHeidelberg}{University of Heidelberg, 68131 Mannheim, Germany}
\newcommand{\instLjubljanaUniLJ}{Faculty of Mathematics and Physics, University of Ljubljana, 1000 Ljubljana, Slovenia}
\newcommand{\instLouisville}{University of Louisville, Louisville, Kentucky 40292, U.S.A.}
\newcommand{\instMalaya}{National Centre for Particle Physics, University Malaya, 50603 Kuala Lumpur, Malaysia}
\newcommand{\instLjubljanaUM}{University of Maribor, 2000 Maribor, Slovenia}
\newcommand{\instMelbourne}{School of Physics, University of Melbourne, Victoria 3010, Australia}
\newcommand{\instMississippi}{University of Mississippi, University, Mississippi 38677, U.S.A.}
\newcommand{\instUOM}{University of Miyazaki, Miyazaki 889-2192, Japan}
\newcommand{\instNovaGorica}{University of Nova Gorica, 5000 Nova Gorica, Slovenia}
\newcommand{\instPittsburgh}{University of Pittsburgh, Pittsburgh, Pennsylvania 15260, U.S.A.}
\newcommand{\instUSTC}{University of Science and Technology of China, Hefei 230026, China}
\newcommand{\instSAlabama}{University of South Alabama, Mobile, Alabama 36688, U.S.A.}
\newcommand{\instSCarolina}{University of South Carolina, Columbia, South Carolina 29208, U.S.A.}
\newcommand{\instSydney}{School of Physics, University of Sydney, New South Wales 2006, Australia}
\newcommand{\instTabuk}{Department of Physics, Faculty of Science, University of Tabuk, Tabuk 71451, Saudi Arabia}
\newcommand{\instUTokyo}{Department of Physics, University of Tokyo, Tokyo 113-0033, Japan}
\newcommand{\instIPMU}{Kavli Institute for the Physics and Mathematics of the Universe (WPI), University of Tokyo, Kashiwa 277-8583, Japan}
\newcommand{\instVictoria}{University of Victoria, Victoria, British Columbia, V8W 3P6, Canada}
\newcommand{\instVPI}{Virginia Polytechnic Institute and State University, Blacksburg, Virginia 24061, U.S.A.}
\newcommand{\instWayneState}{Wayne State University, Detroit, Michigan 48202, U.S.A.}
\newcommand{\instYamagata}{Yamagata University, Yamagata 990-8560, Japan}
\newcommand{\instYerevan}{Alikhanyan National Science Laboratory, Yerevan 0036, Armenia}
\newcommand{\instYonsei}{Yonsei University, Seoul 03722, South Korea}
%%%\affiliation{\instSinica}
\affiliation{\instCPPM}
\affiliation{\instBeihang}
%%%\affiliation{\instBUAP}
\affiliation{\instBNL}
\affiliation{\instBINP}
\affiliation{\instCMU}
\affiliation{\instCinvestavIPN}
\affiliation{\instPrague}
\affiliation{\instChiangMai}
\affiliation{\instChiba}
\affiliation{\instChonnam}
\affiliation{\instConacyt}
\affiliation{\instDESY}
\affiliation{\instDuke}
\affiliation{\instITAR}
%%%\affiliation{\instENEA}
\affiliation{\instEri}
\affiliation{\instJuelich}
\affiliation{\instFuJen}
\affiliation{\instFudan}
\affiliation{\instGoettingen}
\affiliation{\instGifu}
\affiliation{\instSOKENDAI}
\affiliation{\instGyeongsang}
\affiliation{\instHanyang}
\affiliation{\instKEK}
\affiliation{\instJPARC}
\affiliation{\instHSE}
\affiliation{\instIISER}
\affiliation{\instIITBhubaneswar}
\affiliation{\instIITGuwahati}
\affiliation{\instIITHyderabad}
\affiliation{\instIITMadras}
\affiliation{\instIndiana}
\affiliation{\instIHEPRussia}
\affiliation{\instHEPHYVienna}
\affiliation{\instIHEPChina}
%%%\affiliation{\instChennai}
\affiliation{\instIPP}
\affiliation{\instIOP}
\affiliation{\instIFIC}
\affiliation{\instFrascati}
\affiliation{\instNapoliINFN}
\affiliation{\instPadovaINFN}
\affiliation{\instPerugiaINFN}
\affiliation{\instPisaINFN}
\affiliation{\instRomaINFN}
\affiliation{\instRomaTreINFN}
\affiliation{\instTorinoINFN}
\affiliation{\instTriesteINFN}
\affiliation{\instJAEA}
\affiliation{\instMainz}
\affiliation{\instGiessen}
\affiliation{\instKarlsruhe}
%%%\affiliation{\instKennesaw}
\affiliation{\instKitasato}
\affiliation{\instKISTI}
\affiliation{\instKorea}
\affiliation{\instKSU}
%%%\affiliation{\instKyotoU}
\affiliation{\instKyungpook}
\affiliation{\instLPI}
\affiliation{\instLNNU}
\affiliation{\instLMU}
\affiliation{\instLuther}
\affiliation{\instMNITJaipur}
\affiliation{\instMPP}
\affiliation{\instMPGHLL}
\affiliation{\instMcGill}
%%%\affiliation{\instMETU}
\affiliation{\instMEPhI}
\affiliation{\instNagoya}
\affiliation{\instNagoyaKMI}
\affiliation{\instNagoyaIAR}
\affiliation{\instNaraWu}
%%%\affiliation{\instUNAM}
\affiliation{\instNTUTaiwan}
\affiliation{\instNUUTaiwan}
\affiliation{\instKrakow}
\affiliation{\instNiigata}
\affiliation{\instNSU}
\affiliation{\instOkinawa}
\affiliation{\instOsakaCity}
\affiliation{\instRCNP}
\affiliation{\instPNNL}
\affiliation{\instPanjab}
\affiliation{\instPeking}
\affiliation{\instPanjabPAU}
\affiliation{\instRIKENMSL}
%%%\affiliation{\instRIKEN}
%%%\affiliation{\instXavier}
\affiliation{\instSeoul}
%%%\affiliation{\instShandong}
\affiliation{\instSPU}
\affiliation{\instSoochow}
\affiliation{\instSoongsil}
\affiliation{\instLjubljanaJSI}
\affiliation{\instKyiv}
\affiliation{\instTata}
\affiliation{\instTUM}
%%%\affiliation{\instECUTUM}
\affiliation{\instTelAviv}
\affiliation{\instToho}
\affiliation{\instTohoku}
\affiliation{\instTitech}
\affiliation{\instTokyoMetropolitan}
\affiliation{\instUAS}
\affiliation{\instNapoliUNIV}
\affiliation{\instPadovaUNIV}
\affiliation{\instPerugiaUNIV}
\affiliation{\instPisaUNIV}
\affiliation{\instRomaUNIV}
\affiliation{\instRomaTreUNIV}
\affiliation{\instTorinoUNIV}
\affiliation{\instTriesteUNIV}
\affiliation{\instMontreal}
\affiliation{\instIJCLab}
\affiliation{\instIPHC}
\affiliation{\instAdelaide}
\affiliation{\instBonn}
\affiliation{\instUBC}
\affiliation{\instCincinnati}
\affiliation{\instFlorida}
%%%\affiliation{\instHamburg}
\affiliation{\instHawaii}
\affiliation{\instHeidelberg}
\affiliation{\instLjubljanaUniLJ}
\affiliation{\instLouisville}
\affiliation{\instMalaya}
\affiliation{\instLjubljanaUM}
\affiliation{\instMelbourne}
\affiliation{\instMississippi}
\affiliation{\instUOM}
%%%\affiliation{\instNovaGorica}
\affiliation{\instPittsburgh}
\affiliation{\instUSTC}
\affiliation{\instSAlabama}
\affiliation{\instSCarolina}
\affiliation{\instSydney}
%%%\affiliation{\instTabuk}
\affiliation{\instUTokyo}
\affiliation{\instIPMU}
\affiliation{\instVictoria}
\affiliation{\instVPI}
\affiliation{\instWayneState}
\affiliation{\instYamagata}
\affiliation{\instYerevan}
\affiliation{\instYonsei}
  \author{F.~Abudin{\'e}n}\affiliation{\instTriesteINFN} % 2250
  \author{I.~Adachi}\affiliation{\instKEK}\affiliation{\instSOKENDAI} % 2590
  \author{R.~Adak}\affiliation{\instFudan} % 6743
  \author{K.~Adamczyk}\affiliation{\instKrakow} % 2239
  \author{P.~Ahlburg}\affiliation{\instBonn} % 2367
  \author{J.~K.~Ahn}\affiliation{\instKorea} % 7423
  \author{H.~Aihara}\affiliation{\instUTokyo} % 2223
  \author{N.~Akopov}\affiliation{\instYerevan} % 9443
  \author{A.~Aloisio}\affiliation{\instNapoliUNIV}\affiliation{\instNapoliINFN} % 2194
  \author{F.~Ameli}\affiliation{\instRomaINFN} % 4683
% ...
% 2543
\collaboration{Belle II Collaboration}

%% file: introduction.tex
% File info
%   FILE:       introductions.tex
%   FILE OWNER: William Sutcliffe
%   CONTENT:    Intro to FEI, including description of FR and link to FEI pubs.

\section{Introduction}
\label{sec:introduction}

The Belle II experiment~\cite{ref:b2tip} is an $e^{+} e^{-}$ collider experiment in Japan, which began its main physics runs in early 2019 and has collected $74$~fb$^{-1}$ of data at a centre-of-mass (CM) energy, $\sqrt{s}$, corresponding to the mass of the $\Upsilon(4S)$ resonance. The clean environment of $e^{+}e^{-}$ collisions together with the unique event topology of Belle II, in which an $\Upsilon(4S)$ meson is produced and subsequently decays in a pair of $B$ mesons, allows a wide range of physics measurements to be performed that are difficult or impossible at hadron colliders. In particular, measurements in which there is missing energy, which includes semileptonic decays with missing neutrinos, can benefit substantially from the additional constraints provided by the collision environment of Belle II. This includes the measurement of the ratio of branching fractions, $R(D^{*}) = \mathcal{B}(B\rightarrow D^{(*)} \tau \nu) / \mathcal{B}(B\rightarrow D^{(*)} \ell \nu)$, inclusive determinations of the CKM matrix elements $|V_{ub}|$ and $|V_{cb}|$ from $B \rightarrow X_{u/c}\ell \nu$ decays and searches for the rare decay $B \rightarrow K^{*} \nu \bar{\nu}$.

Full Event Interpretation~\cite{Keck:2018lcd} is an algorithm for tag-side $B$ meson reconstruction at Belle II. The algorithm utilises a hierarchical reconstruction of exclusive decay chains of $B$ mesons, with multivariate classifiers utilised to identify each unique sub-decay channel. Given the large number of decay chains reconstructed and multivariate classifiers employed, there can be significant differences between the tag-side reconstruction efficiency in simulation and data. In order to correct for this, a calibration can be performed by measuring a decay with a well known branching fraction and sufficient available statistics after selection. A suitable choice, given the current Belle II dataset, is inclusive $B \rightarrow X \ell \nu$ decays due to their substantial branching fraction of $\sim$$20$\%.

%% file: detector_and_simulation.tex
% File info
%   FILE:       detector_and_simulation.tex
%   FILE OWNER: Peter Lewis
%   CONTENT:    Standard Belle II description. 

\section{Detector and simulation}
\label{sec:detector_and_simulation}

The Belle~II detector~\cite{Abe:2010sj, ref:b2tip} operates at the SuperKEKB asymmetric-energy  electron-positron collider~\cite{superkekb}, located at the KEK laboratory in Tsukuba, Japan. 
The detector consists of several nested detector subsystems arranged around the beam pipe in a cylindrical geometry. 

The innermost subsystem is the vertex detector, %(VXD), 
which includes two layers of silicon pixel detectors and four outer layers of silicon strip detectors.
Currently, the second pixel layer is installed in only a small part of the solid angle, while the remaining vertex detector %VXD 
layers are fully installed. 
Most of the tracking volume consists of a helium- and ethane-based small-cell drift chamber. %(CDC).

Outside the drift chamber, %CDC, 
a Cherenkov-light imaging and time-of-propagation detector %(TOP) 
provides charged-particle identification in the barrel region. In the forward endcap, this function is provided by a proximity-focusing, ring-imaging Cherenkov detector with an aerogel radiator. %(ARICH).
Further out is an electromagnetic calorimeter, %(ECL) 
consisting of a barrel and two endcap sections made of CsI(Tl) crystals. A uniform 1.5~T magnetic field is provided by a superconducting solenoid situated outside the calorimeter. %ECL.
Multiple layers of scintillators and resistive plate chambers, located between the magnetic flux-return iron plates, constitute the $K_L$ and muon %(KLM) 
identification system.

%{\color{red} Describe the current lepton ID algorithms: using only ECL and KLM, or also dE/dX, TOP, ARICH?}

The data used in this analysis were collected at a CM energy, $\sqrt{s}$, of 10.58~GeV, corresponding to the mass of the $\Upsilon$(4S) resonance. 
The energies of the electron and positron beams are $7\gev$ and $4\gev$, respectively, resulting in a boost of $\beta\gamma = 0.28$ of the CM frame relative to the lab frame. 
The integrated luminosity of the data is \lumi. In addition, a smaller sample of $3.23$ fb$^{-1}$ off-resonance data was collected at a CM energy of $10.52$~GeV.

The analysis utilises several samples of simulated events. These include a sample of $e^+ e^-\to (\Upsilon(4S) \to B\bar B)$ with generic $B$-meson decays, generated with EvtGen~\cite{Lange:2001uf}, and corresponding to an integrated luminosity of 100~\ifb. A 100~\ifb sample of continuum $e^+e^-\to q\bar q~ (q = u, d, s, c)$ is simulated with KKMC~\cite{ref:kkmc} interfaced with PYTHIA~\cite{Sjostrand:2007gs}. All data samples were analyzed (and, for Monte Carlo (MC) events, generated and simulated) in the basf2~\cite{basf2} framework.

%% file: algorithm.tex
% File info
%   FILE:       algorithm.tex
%   FILE OWNER: William Sutcliffe
%   CONTENT:    Description of FEI algorithm

\section{The Algorithm}
\label{sec:algorithm}

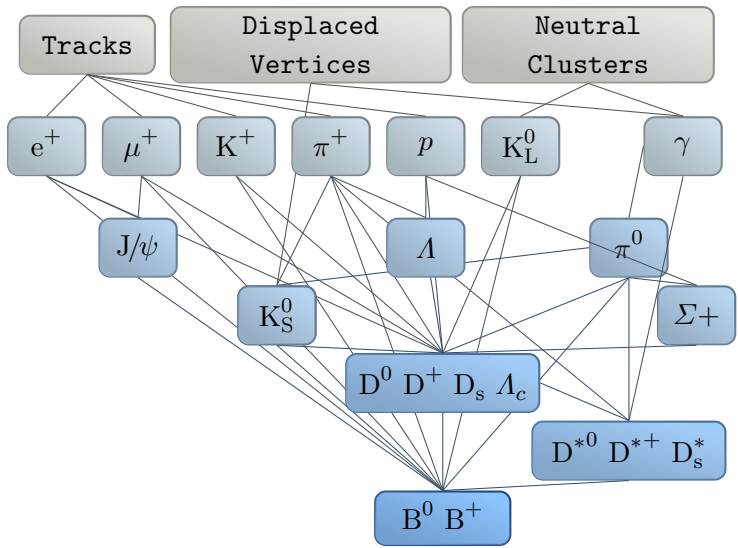
\begin{figure}[h]
\begin{center}
\begin{tikzpicture}[%
scale=0.9,
    detector/.style={
      rectangle,
      draw=Tgray,
      thick,
      font=\small,
      fill=Tgraylight2,
      postaction={path fading=north, fading angle=-45, fill=Tgraylight},
      text width=10em,
      align=center,
      rounded corners,
      minimum height=2em
    },  
    fsp/.style={
      rectangle,
      draw=Tblue!25!Tgray,
      thick,
      font=\small,
      fill=Tbluelighter!25!Tgraylight2,
      postaction={path fading=north, fading angle=-45, fill=Tbluelight!25!Tgraylight},
      text width=2em,
      align=center,
      rounded corners,
      minimum height=2em
    },  
    greenfsp/.style={
      rectangle,
      draw=Tblue!50!Tgray,
      thick,
      font=\small,
      fill=Tbluelighter!50!Tgraylight2,
      postaction={path fading=north, fading angle=-45, fill=Tbluelight!50!Tgraylight},
      text width=2em,
      align=center,
      rounded corners,
      minimum height=2em
    },  
    dstage/.style={
      rectangle,
      draw=Tblue!75!Tgray,
      thick,
      font=\small,
      fill=Tbluelighter!75!Tgraylight2,
      postaction={path fading=north, fading angle=-45, fill=Tbluelight!75!Tgraylight},
      text width=6em,
      align=center,
      rounded corners,
      minimum height=2em
    },  
    bstage/.style={
      rectangle,
      draw=Tbluedark,
      thick,
      font=\small,
      fill=Tbluelighter,
      postaction={path fading=north, fading angle=-45, fill=Tbluelight},
      text width=4em,
      align=center,
      rounded corners,
      minimum height=1em
    },  
  ]

\draw (1.0,1) node[detector, text width=4em] (Tr) {\texttt{Tracks}};
\draw (4.3,1) node[detector,text width=9em] (Vo) {\texttt{Displaced Vertices}};
\draw (8.4,1) node[detector,text width=8em] (Cl) {\texttt{Neutral Clusters}};

\draw (9,-2) node[greenfsp] (Pio) {$\Ppizero$};

\draw (7.4,-0.5) node[fsp] (Kaol) {$\PKlong$};
\draw[Tgraydark] (Cl.south) -> (Kaol.north);

\draw (3.8,-3) node[greenfsp] (Kao) {$\PKshort$};
\draw[Tgraydark] (Vo.south) -> (Kao.north);
\draw (4.6,-0.5) node[fsp] (Pi) {$\Ppiplus$};
\draw[Tbluedark!25!Tgraydark] (Pi.south) -> (Kao.north);
\draw[Tbluedark!50!Tgraydark] (Pio.west) -> (Kao.north);

\draw (0.4,-0.5) node[fsp] (El) {$\Pep$};
\draw (1.8,-0.5) node[fsp] (Mu) {$\APmuon$};
\draw (3.2,-0.5) node[fsp] (Ka) {$\PKplus$};
\draw (6,-0.5) node[fsp] (Pr) {$p$};
\draw[Tgraydark] (Tr.south) -> (El.north);
\draw[Tgraydark] (Tr.south) -> (Mu.north);
\draw[Tgraydark] (Tr.south) -> (Pi.north);
\draw[Tgraydark] (Tr.south) -> (Ka.north);
\draw[Tgraydark] (Tr.south) -> (Pr.north);

\draw (10,-3) node[greenfsp] (Sig) {$\Sigma+$};
\draw[Tbluedark!25!Tgraydark] (Pr.south) -> (Sig.north);
\draw[Tbluedark!50!Tgraydark] (Pio.south) -> (Sig.north);

\draw (9.8,-0.5) node[fsp] (Ga) {$\Pgamma$};

\draw[Tgraydark] (Vo.south) -> (Ga.north);
\draw[Tgraydark] (Cl.south) -> (Ga.north);
\draw[Tbluedark!25!Tgraydark] (Ga.west) -> (Pio.north);

\draw (9.0,-5) node[dstage] (DStar) {$\PDst^0 \ \PDst^+ \ \PDsstar$};

\draw[Tbluedark!25!Tgraydark] (Pi.south) -> (DStar.north);
\draw[Tbluedark!50!Tgraydark] (Pio.south) -> (DStar.north);
\draw[Tbluedark!25!Tgraydark] (Ga.south) -> (DStar.north);

\draw (6.25,-6) node[bstage] (B) {$\PBzero\ \PBplus$};
\draw[Tbluedark!25!Tgraydark] (Ka.south) -> (B.north);
\draw[Tbluedark!25!Tgraydark] (Pi.south) -> (B.north);
\draw[Tbluedark!25!Tgraydark] (Kaol.south) -> (B.north);
\draw[Tbluedark!50!Tgraydark] (Pio.south) -> (B.north);
\draw[Tbluedark!50!Tgraydark] (Kao.south) -> (B.north);

\draw[Tbluedark!25!Tgraydark] (El.south) -> (B.north);
\draw[Tbluedark!25!Tgraydark] (Mu.south) -> (B.north);

\draw[Tbluedark!50!Tgraydark] (DStar.south) -> (B.north);

\draw (6.25,-4) node[dstage] (D) {$\PDzero\ \PDplus\ \PDs$ $\Lambda_{c}$};
\draw[Tbluedark!25!Tgraydark] (Ka.south) -> (D.north);
\draw[Tbluedark!25!Tgraydark] (Pi.south) -> (D.north);
\draw[Tbluedark!25!Tgraydark] (Kaol.south) -> (D.north);
\draw[Tbluedark!25!Tgraydark] (El.south) -> (D.north);
\draw[Tbluedark!25!Tgraydark] (Mu.south) -> (D.north);
\draw[Tbluedark!50!Tgraydark] (Kao.south) -> (D.north);
\draw[Tbluedark!50!Tgraydark] (Pio.south) -> (D.north);
\draw[Tbluedark!50!Tgraydark] (Sig.south) -> (D.north);
\draw[Tbluedark!50!Tgraydark] (Pr.south) -> (D.north);
\draw[Tbluedark!50!Tgraydark] (D.east) -> (DStar.north);
\draw[Tbluedark!50!Tgraydark] (D.south) -> (B.north);

\draw (1.75,-2) node[greenfsp] (J) {$\PJpsi$};
\draw[Tbluedark!25!Tgraydark] (El.south) -> (J.north);
\draw[Tbluedark!25!Tgraydark] (Mu.south) -> (J.north);
\draw[Tbluedark!50!Tgraydark] (J.south) -> (B.north);

\draw (6,-2) node[greenfsp] (Lam) {$\Lambda$};
\draw[Tbluedark!25!Tgraydark] (Pr.south) -> (Lam.north);
\draw[Tbluedark!25!Tgraydark] (Pi.south) -> (Lam.north);
\draw[Tbluedark!50!Tgraydark] (Lam) -> (D.north);

\draw (3.8,-3) node[greenfsp] (Kao) {$\PKshort$};
\end{tikzpicture}
\end{center}
\caption{The stages of reconstruction employed by Full Event Interpretation.}
\label{fig:FEIstages}
\end{figure}

The Full Event Interpretation employs a hierarchical reconstruction of exclusive $B$ meson decay chains, in which each unique decay channel of a particle has its own designated multivariate classifier. The algorithm utilises several stages of reconstruction, which are shown in Fig.~\ref{fig:FEIstages}. The algorithm starts by selecting candidates for stable particles, which include muons, electrons, pions, kaons, protons and photons, from tracks and EM clusters in the event. Subsequently, the algorithm carries out several stages of reconstruction of intermediate particles such as $\pi^{0}$, $K^{0}_{S}$, $\jpsi$, $D$ and $D^{*}$ mesons and, in addition, $\Sigma$, $\Lambda$ and $\Lambda_{c}$ baryons. The addition of baryonic modes is a recent extension of the algorithm. Intermediate particles are reconstructed in specific decay modes from a combination of stable and other intermediate particle candidates. The final stage of the algorithm reconstructs the $B^{+}$ and $B^{0}$ mesons in 36 (8) and 31 (8)  hadronic (semileptonic) modes.

\begin{figure}[h!]
\begin{center}
\minipage{0.45\textwidth}
	\includegraphics[width=7cm]{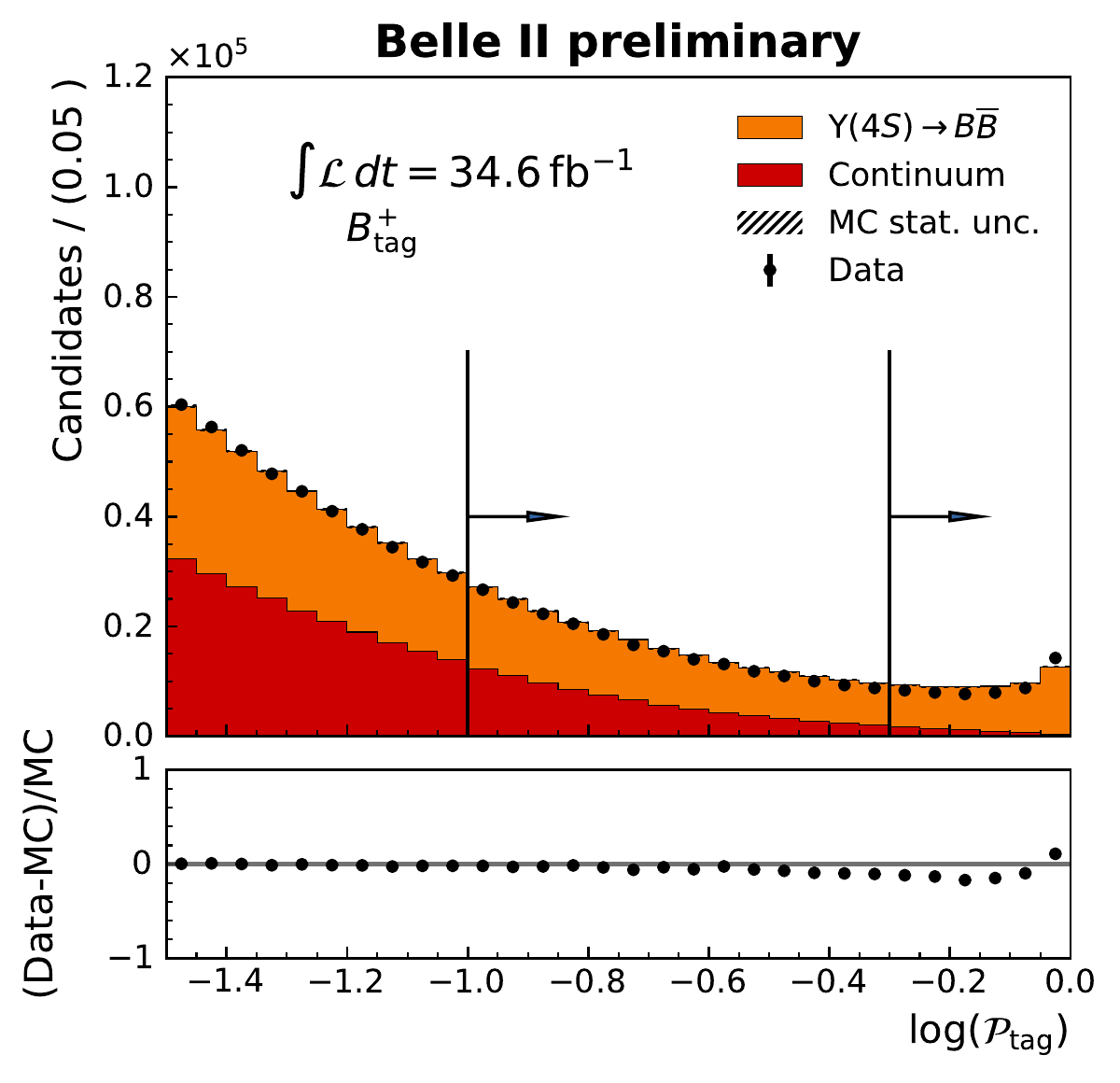} \put(-210,180){(a)}
\endminipage\hfill
\minipage{0.275\textwidth}
\includegraphics[width=5cm]{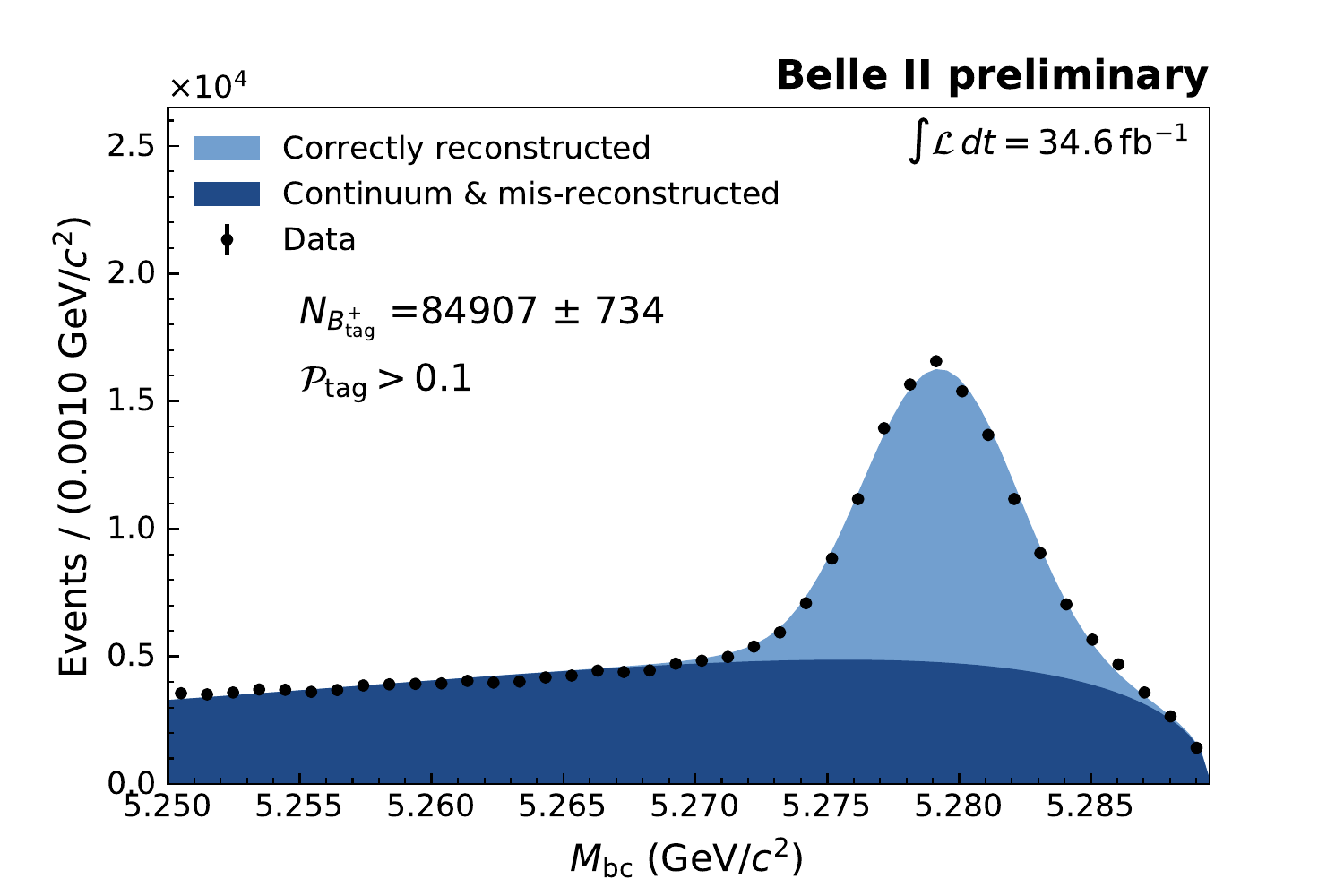} \put(-150,90){(b)}\\
\includegraphics[width=5cm]{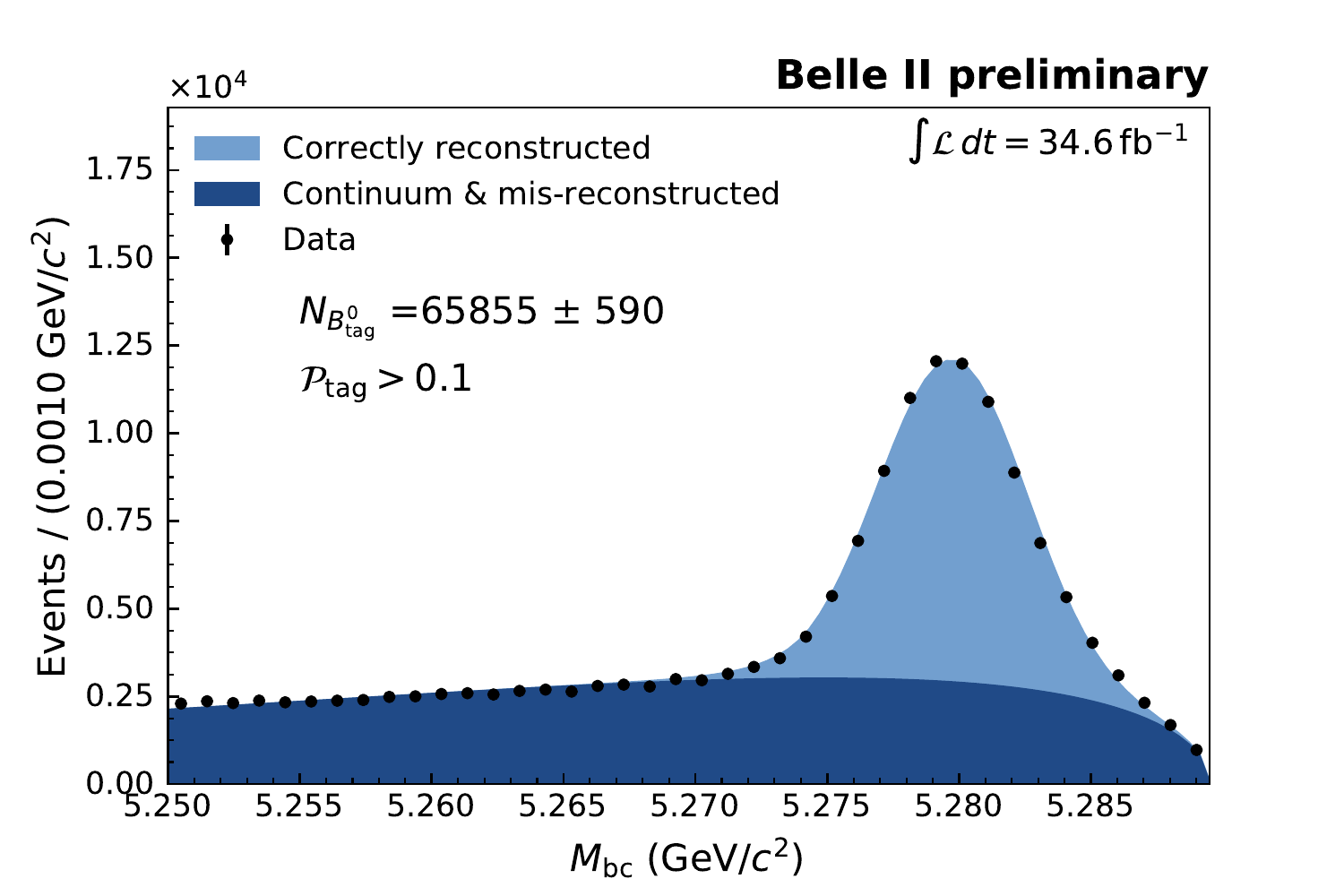} 
\endminipage\hfill
\minipage{0.275\textwidth}\hfill
 \includegraphics[width=5cm]{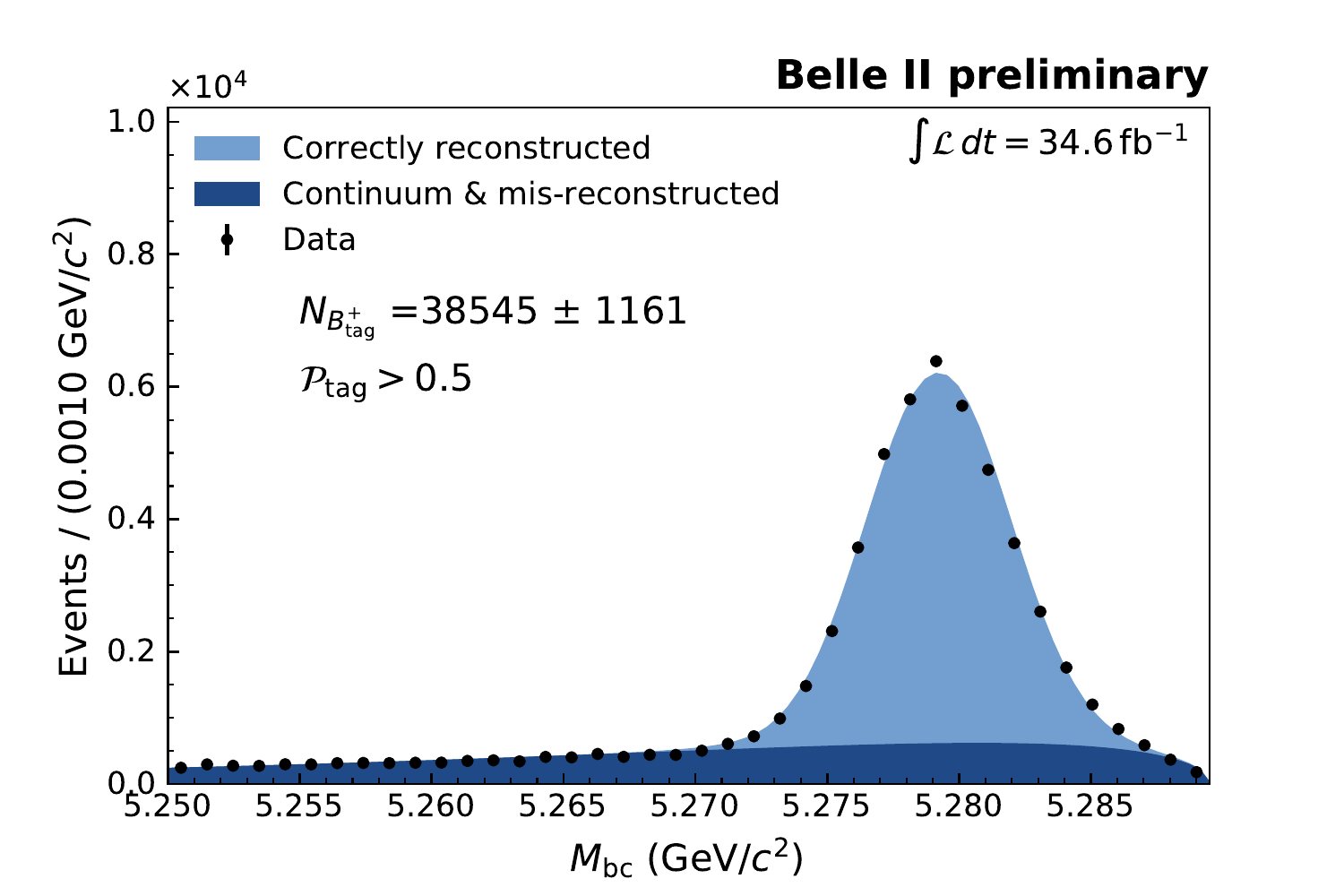} \\
 \includegraphics[width=5cm]{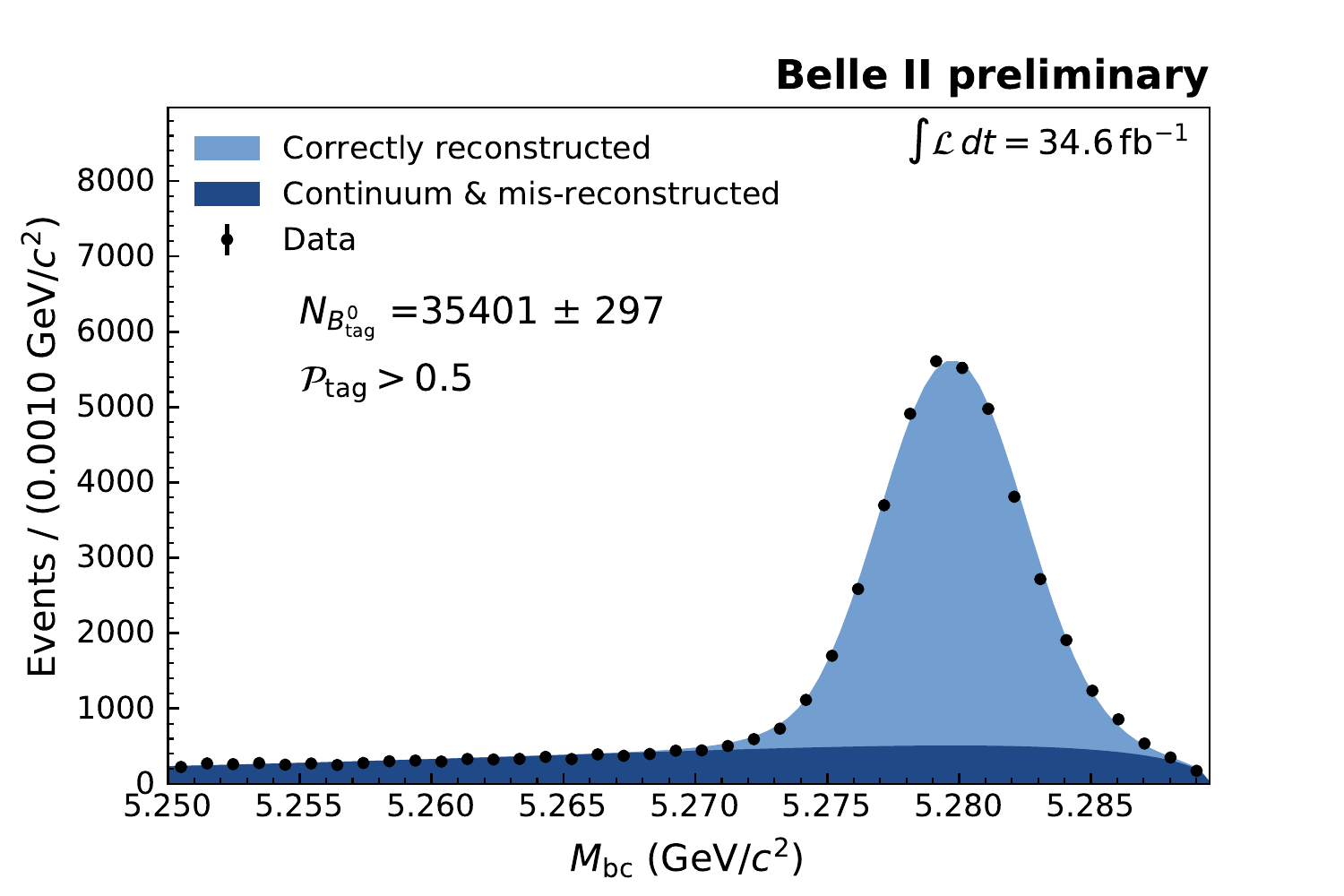} 
\endminipage
\end{center}
	\caption{(a) Comparison of the distribution of $\log \mathcal{P_{\rm tag}}$ in early Belle II data to the shape expectation from simulation. Here, $\log \mathcal{P_{\rm tag}}$ is the logarithm of the tag-side $B^{+}$ meson classifier output, $ \mathcal{P_{\rm tag}}$. Reference selection criteria of $\mathcal{P_{\rm tag}} > 0.1$ and $\mathcal{P_{\rm tag}} > 0.5$ are illustrated. (b) Fits to the beam-constrained-mass, $M_{\rm bc}$, distribution of reconstructed $B^{+}$ (top) and $B^{0}$ (bottom) tag-side $B$ mesons in data. A looser selection criteria of $\mathcal{P_{\rm tag}} > 0.1$ (left) and a tighter selection criteria of $\mathcal{P_{\rm tag}} > 0.5$ (right) are applied on the $B$ meson classifier $\mathcal{P_{\rm tag}}$ to select samples with different levels of purity.}
\label{fig:fits}
\end{figure}

Each stage consists of pre-reconstruction and post-reconstruction steps. In the pre-reconstruction step, candidates for particles are reconstructed, an inital pre-selection is applied and a best candidate selection is made on a discriminating variable. Subsequently, in the post-reconstruction step, vertex fits are performed where applicable, pre-trained classifiers are applied and a best-candidate selection is made on the classifier output. Classifiers for stable particles utilise kinematic and particle identification information as features; meanwhile, intermediate and $B$ classifiers utilise the kinematic information from all daughters, daughter classifier outputs and information from vertex fits as features.

The algorithm requires a training procedure, in which all of the particle classifiers are trained. For the calibration studies performed here, the training was performed on simulated $\Upsilon(4S) \rightarrow B\bar{B}$ events corresponding to an integrated luminosity of $100$~fb$^{-1}$. The training of the algorithm utilises an equivalent reconstruction procedure to produce training datasets for each particle decay channel classifier.

Subsequently, the tag-side $B$ classifier, $\mathcal{P_{\rm tag}}$, can be used to select a pure sample of correctly reconstructed tag-side $B$ mesons. This is demonstrated in Fig.~\ref{fig:fits}, which shows fits to the beam constrained mass distribution, $M_{\rm bc} = \sqrt{E^{2}_{\rm beam} - (p^{\rm CM}_{\rm tag})^{2}}$, for reconstructed tag-side $B^{0}$ and $B^{+}$ mesons, for selections requiring $\mathcal{P_{\rm tag}}$ to be greater than 0.1 and 0.5. The contribution from correctly reconstructed tag-side $B$ mesons is parametrised by a Crystal Ball function~\cite{Skwarnicki:1986xj}; backgrounds from $e^{+} e^{-} \rightarrow q \bar{q}$ and incorrectly reconstructed $B$ mesons are modelled with an Argus function~\cite{Albrecht:1990am}. By applying a tighter selection on the classifier output, a higher purity sample of tag-side $B$ mesons can be selected with the sacrifice of a lower tag-side efficiency, which is proportional to the yield of correctly reconstructed tag-side $B$ mesons.

%% file: selection.tex
% File info
%   FILE:       performance.tex
%   FILE OWNER: William Sutcliffe
%   CONTENT:    Report on FEI performance studies. Subsections for each separate author/
%               performance study. 

\section{Selection}
\label{sec:selection}

\begin{figure}[h!]
\begin{center}
\includegraphics[width=7.5cm]{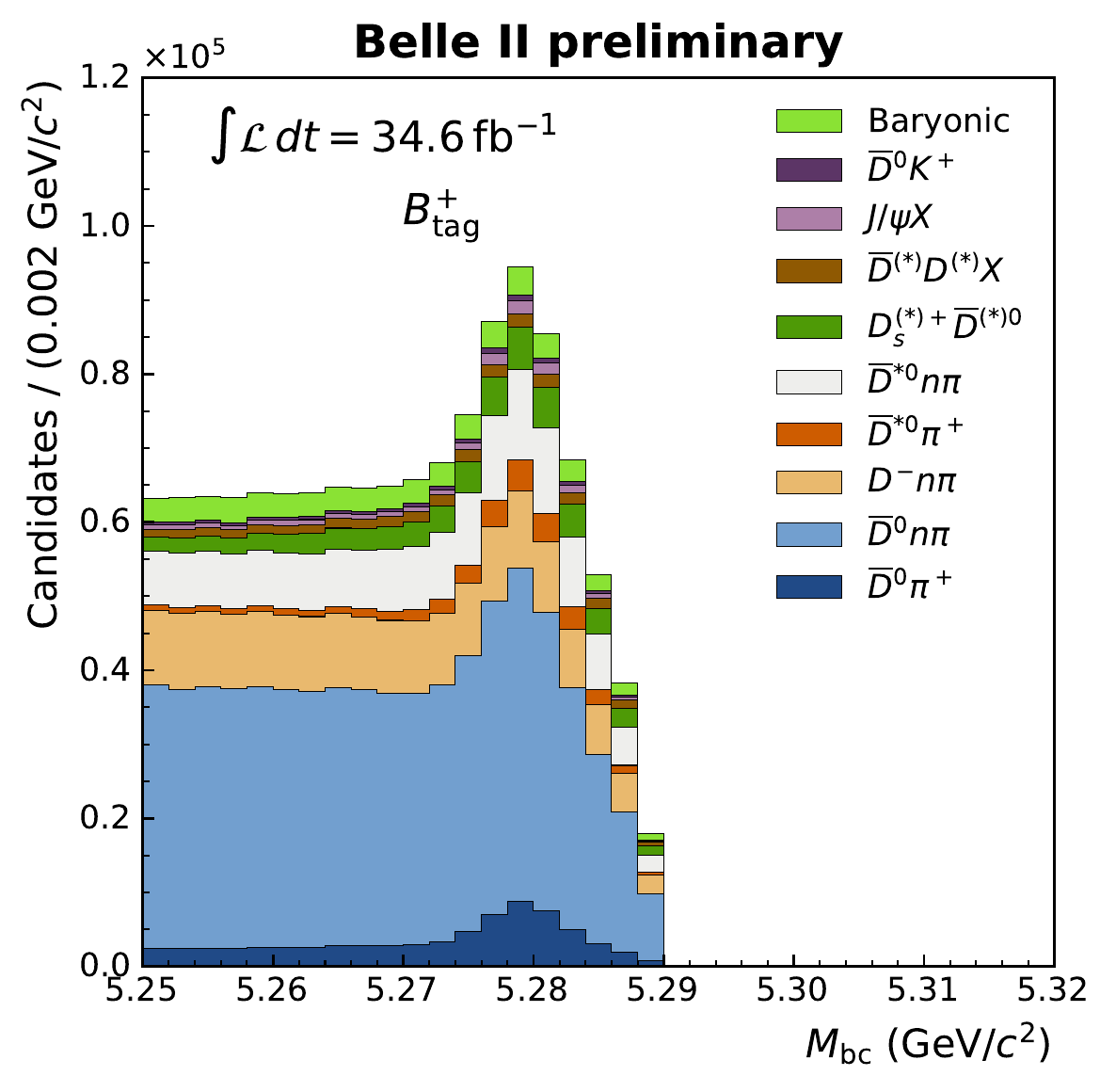} \includegraphics[width=7.5cm]{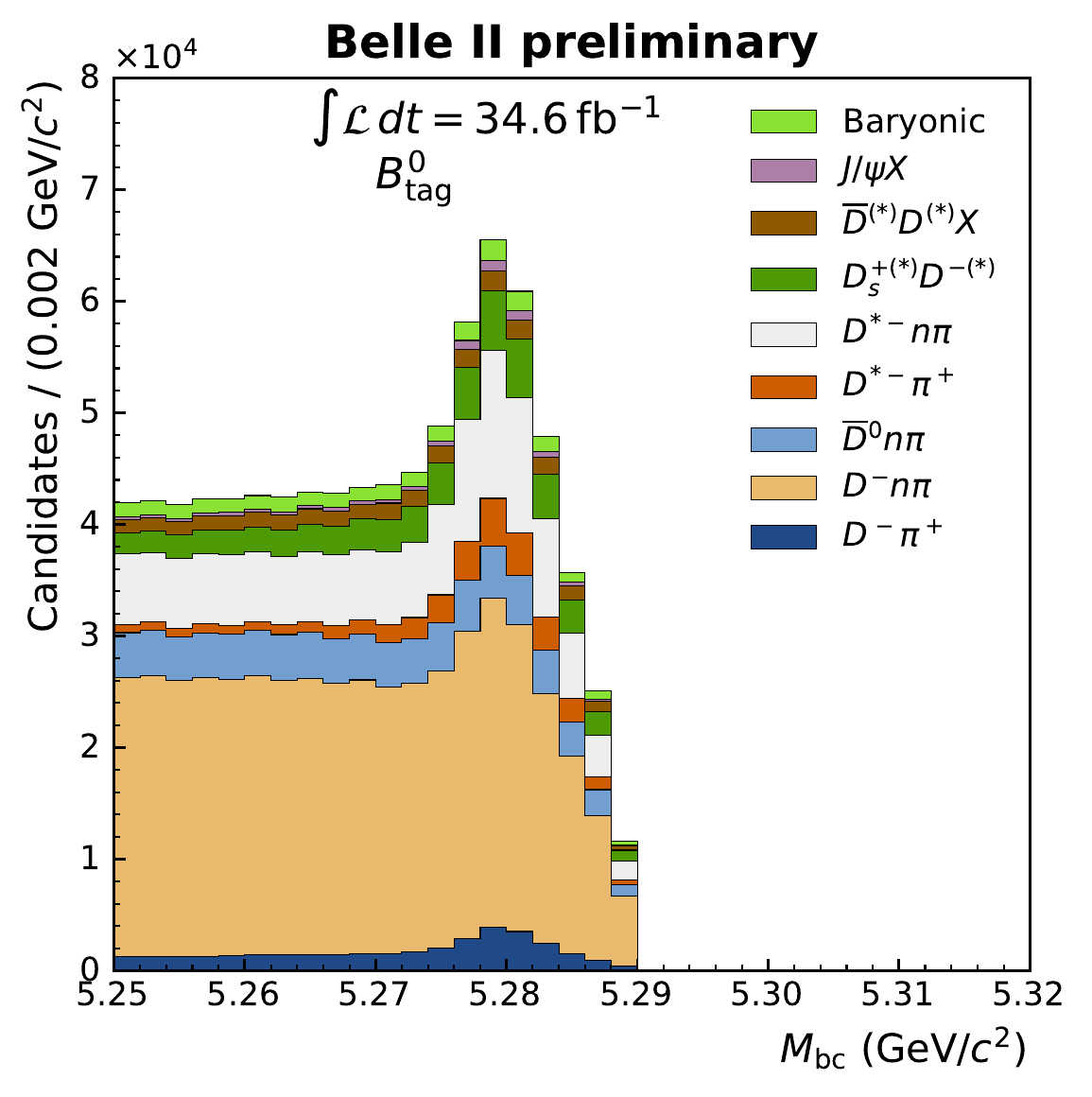} \\ 
	\caption{Contribution of different tag-side decay modes to the $M_{ \rm bc}$ distribution in data for $B^{+}$ (left) and $B^{0}$ (right) parents for $\mathcal{P_{\rm tag}} > 0.01$. Contributions from the newly added baryonic modes can also be seen.}
  \label{fig:mbcmodes}
\end{center}
\end{figure}

The selection process begins by requiring that there is at most one tag-side $B$ meson candidate in each event. This is achieved by selecting the tag-side candidate with the highest tag-side $B$ classifier output, $\mathcal{P_{\rm tag}}$. For correctly reconstructed tags, the beam energy difference, $\Delta E$, should peak around $0$ with some mode-dependent resolution, which is asymmetric with a skew towards lower values for modes containing $\pi^{0} \rightarrow \gamma \gamma$ decays. Therefore, an asymmetric requirement of $-0.15< \Delta E < 0.1$~GeV is placed on the beam energy difference. To reduce background from $e^{+} e^{-} \rightarrow q \bar{q}$ events, a requirement on the event-level-normalised second Fox-Wolfram moment to be less than 0.3 is made. Fig.~\ref{fig:mbcmodes} shows a breakdown of the $M_{bc}$ distribution in data into several categories of tag-side decay mode after the above selection and the loose purity requirement that $\mathcal{P_{\rm tag}} > 0.01$. The dominant tag-side decay mode categories are $D \pi$, $D^{*} \pi$, $D n \pi$ and $D^{*} n \pi$. The recently added baryonic modes result in a small increase in the tag-side efficiency, boosting the number of correctly reconstructed tag-side $B$ mesons by roughly 3\% (2\%) for tag-side $B^{+}$ ($B^{0}$) mesons. The final selection applied to the tag-side candidate, is a requirement that $M_{bc}$ is greater than 5.27~GeV/$c^{2}$, which selects the region containing correctly reconstructed tag-side $B$ mesons as can be seen in Fig.~\ref{fig:fits}.

%The performance of a tag-side reconstruction algorithm can be quantified by a number metrics. The first of these is the tagging efficiency, which is ratio of the number of events containing a tag-side to the total number of $\Upsilon(4S)$ events, $N_{\Upsilon(4S)}$. Of greater importance for a physics analysis is the tag-side efficiency, which is the number of correctly reconstructed tag-side events to $N_{\Upsilon(4S)}$. The last metric is the purity, which is the ratio of the tag-side efficiency to the tagging efficiency. While it is possible to select a sample with higher purity by tightening a selection on the tag-side $B$ meson classifier this will come with a trade off of a lower tag-side efficiency

After the tag-side selection, the signal-side selection is applied. In particular, a lepton is selected with $p^{*}_{\ell} > 1$~GeV/$c$, where $p^{*}_{\ell}$ refers to the momentum of the lepton in the rest frame of the signal-side $B$ meson, which can be determined using the four-momentum of the recoiling tag-side $B$ meson. The distance of closest approach between each track and the interaction point is required to be less than 2 cm along the $z$ direction (parallel to the beams) and less than 0.5 cm in the transverse $r-\phi$ plane. Particle identification information from several sub-detectors, including Cherenkov time of propagation (TOP), Aerogel ring imaging Cherenkov and dedicated muon detectors, is combined into a likelihood for each of electron and muon hypotheses in order to select each lepton species. The selection on $p^{*}_{\ell}$ to be greater than $1$~GeV/$c$ was motivated by the fact that lepton identification performance is found to degrade significantly below $1$~GeV/$c$. 
%The number of correctly reconstructed tag-side $B$ mesons, $N_{B}$, was determined in data by fitting the beam constrained mass $m_{bc}$ ditribution of candidate tag-side $B$ mesons. The $m_{bc}$ shape of correctly reconstucted B mesons was parametrised with a crystal ball, meanwhile, background from mis-reconstructed tag-side $B$ mesons and continuum was modelled using an Argus function. Fig. X shows compares Mbc fits for $B^{+}$ and $B^{0}$ mesons in high an low purity regions. Fig. Y shows the tag-side efficiency against purity for several selections on the $B$ classifiers.  

%% file: calibration.tex
% File info
%   FILE:       calibration.tex
%   FILE OWNER: William Sutcliffe
%   CONTENT:    Report on FEI calibration studies. Subsections for each separate author/
%               calibration study. 

\section{Calibration procedure}
\label{sec:calibration}

The calibration factor is defined as $\epsilon =  N^{\rm Data}_{X\ell\nu} / N^{\rm MC}_{X\ell\nu}$, where the yield of $B \to X\ell\nu$ decays in data, $N^{\rm Data}_{X\ell\nu}$, is determined by fitting the $p^{*}_{\ell}$ distribution and the expected yield, $N^{\rm MC}_{X\ell\nu}$, is determined using MC simulation. 

The fitting procedure maximises a binned likelihood, $\mathcal{L}$, defined by the following equation,
\begin{equation}
\begin{split}
	-2 \log \mathcal{L}  =  -2 \log \prod_{i} {\rm P}(\nu^{\rm obs}_{i} | \nu^{\rm exp}_{i})  + \theta^{T} \Sigma^{-1}_{\theta} \theta^{T}  + (k - k_{\rm constraints})^{T} \Sigma^{-1}_{\rm constraints} (k - k_{\rm constraints})\; ,
\end{split} 
	\label{eq:NLL}
\end{equation}
where the probability to observe $\nu^{\rm obs}_{i}$ events in bin $i$ of $p^{*}_{\ell}$ given that $\nu^{\rm exp}_{i}$ events were expected is ${\rm P}(\nu^{\rm obs}_{i} | \nu^{\rm exp}_{i})$ and is governed by a Poisson distribution. Here, $\nu^{\rm exp}_{i}$, is given by
\begin{equation}
\begin{split}
	\nu^{\rm exp}_{i} = \sum_{j} \nu^{j}   \frac{ p^{j}_{i}(1 + \theta^{j}_{i} )}{  \sum_{k}  p^{j}_{k} (1 + \theta^{j}_{k} ) }\; ,
\end{split} 
\end{equation}
where $p^{j}_{i}$ defines the probability for an event of process type $j$ to have a reconstructed value of $p^{*}_{\ell}$ in bin $i$. The nuisance parameters, $\theta^{j}_{i}$, account for both MC template statistics and additional systematic effects. The associated bin-to-bin correlations arising from systematic uncertainties are accounted for in the covariance matrix, $\Sigma_{\theta}$. 

\begin{figure}[h!]
\begin{center}
	\includegraphics[width=7cm]{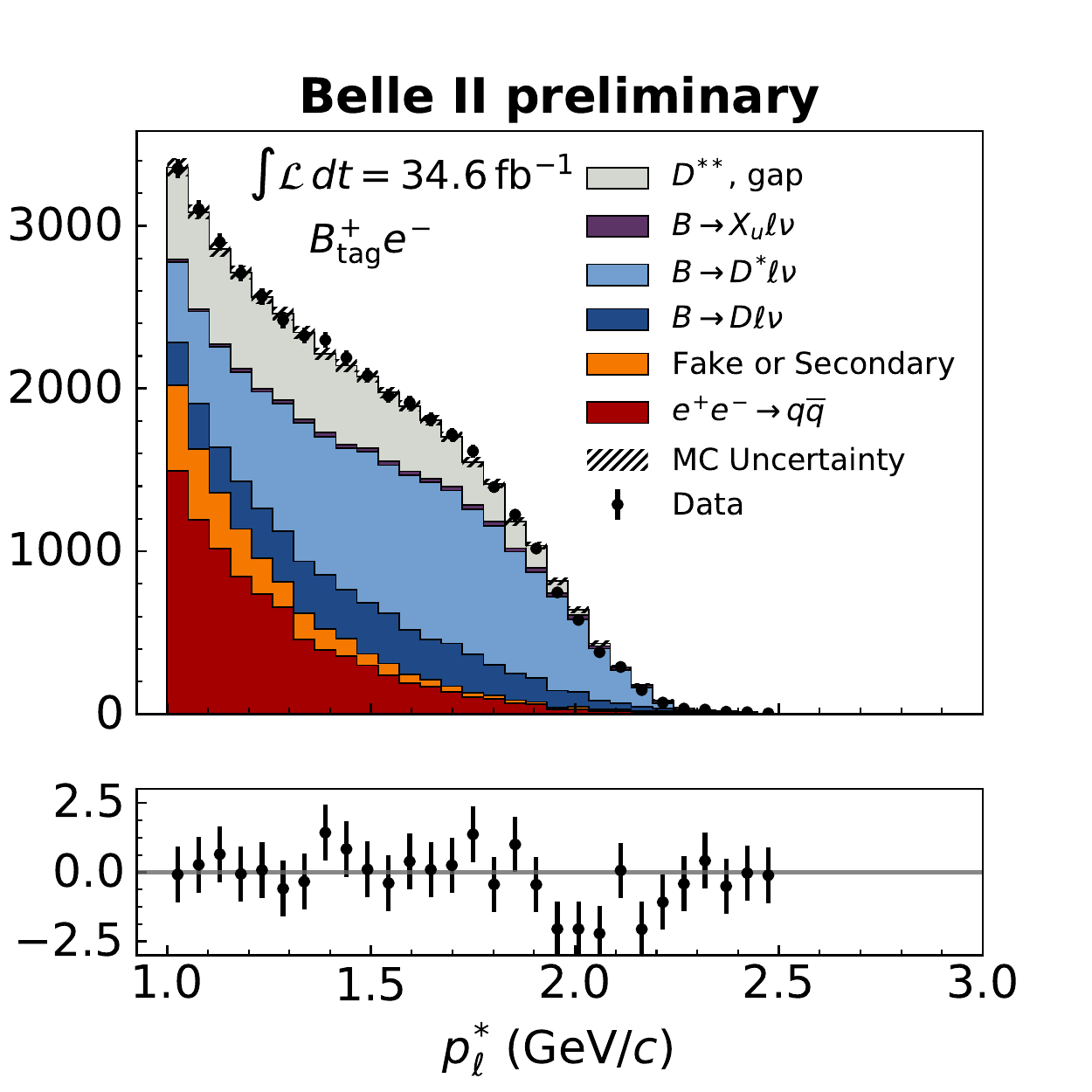} \includegraphics[width=7cm]{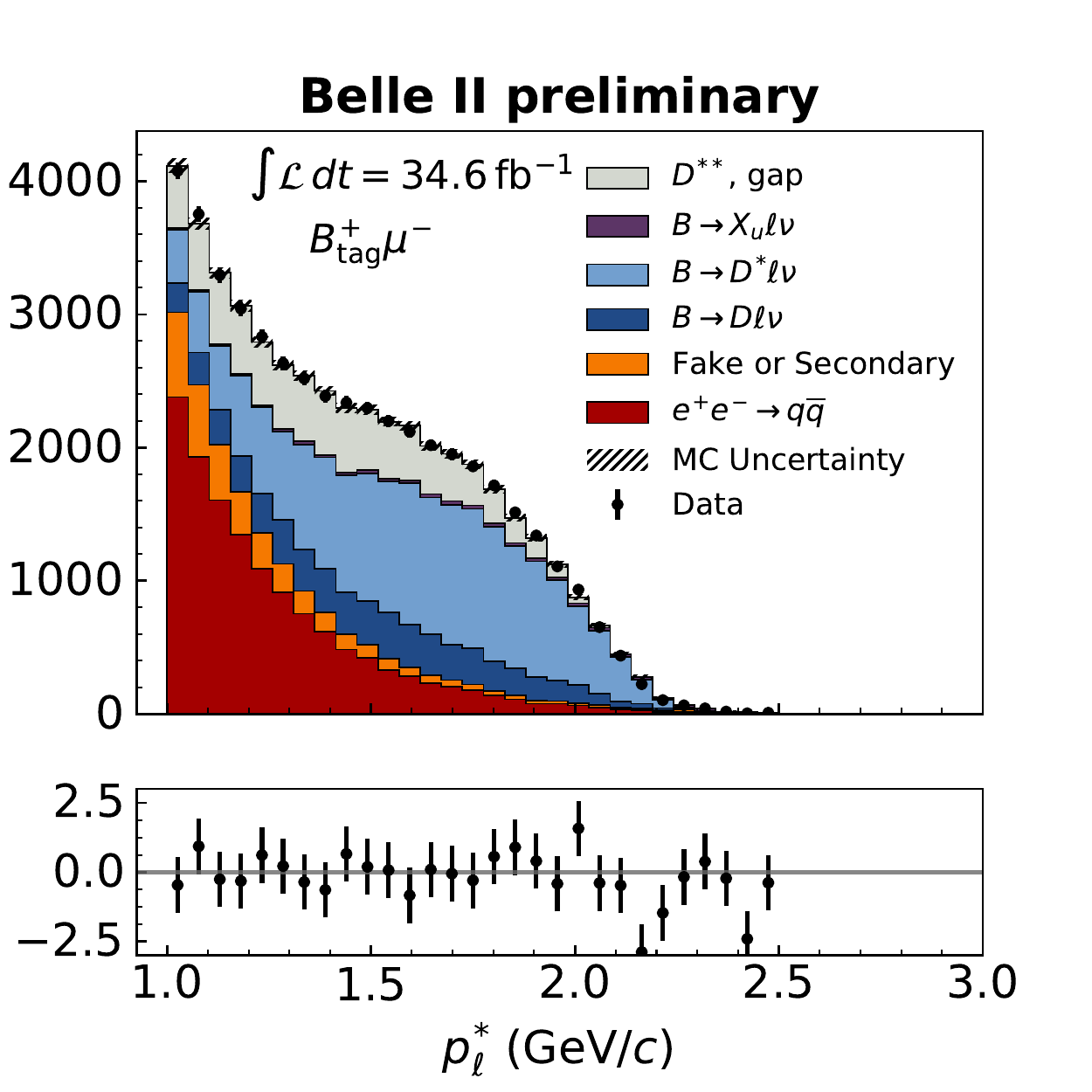} \\
 \includegraphics[width=7cm]{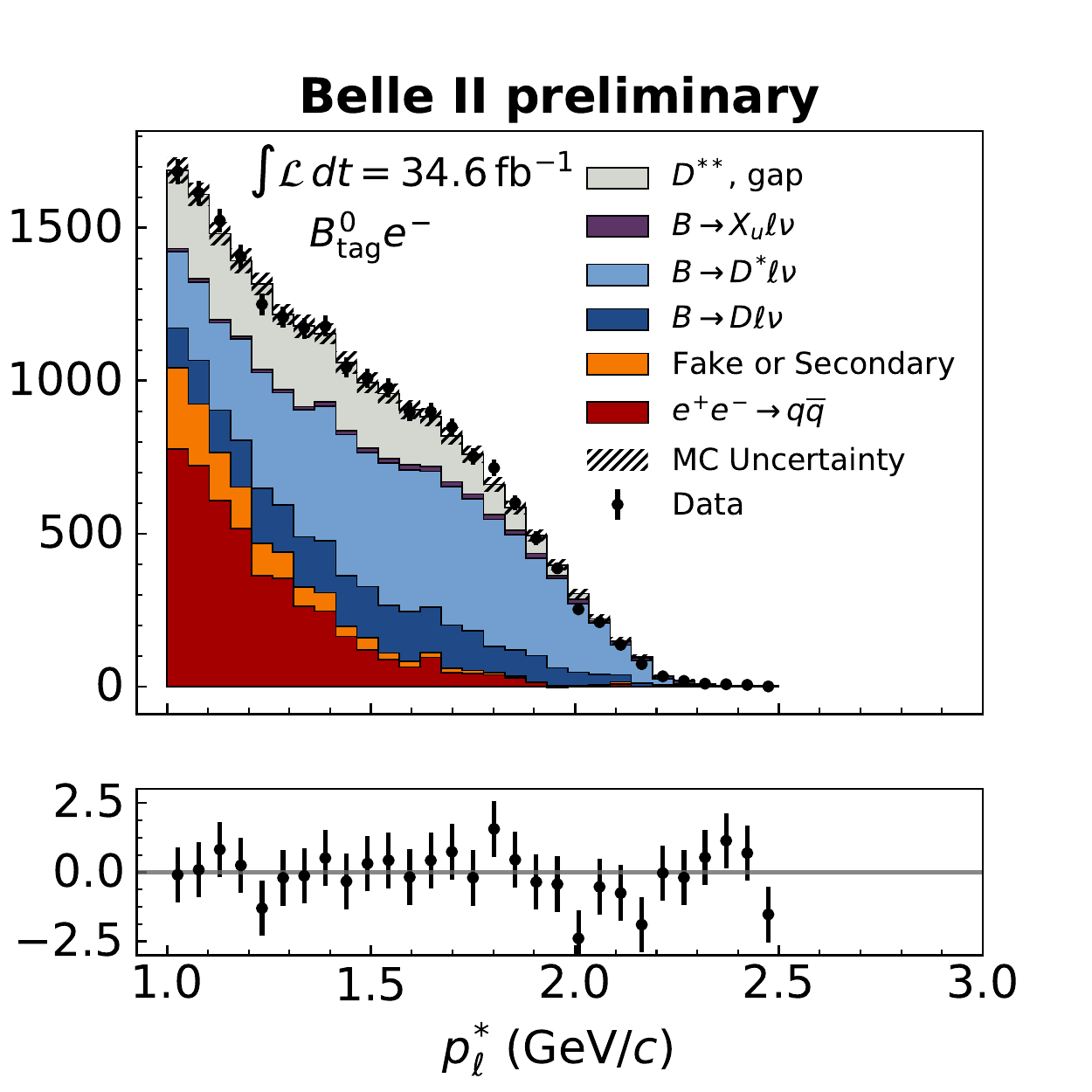} \includegraphics[width=7cm]{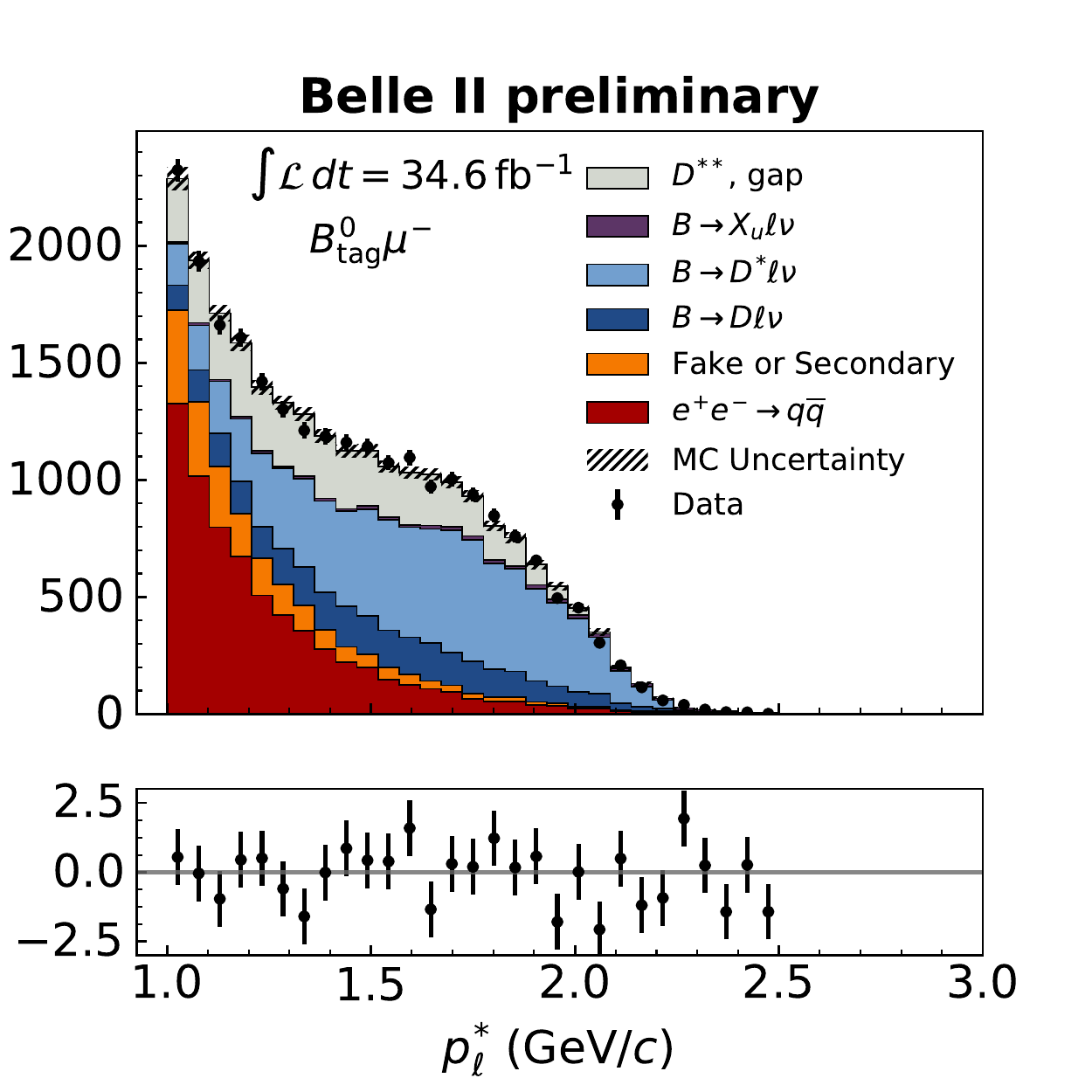}
	\caption{Fits to $p^{*}_{\ell}$ in data for charged (top) and neutral (bottom) tag-side $B$ mesons combined either with electron (left) or muon (right) signal-side $B \rightarrow X\ell \nu$ decays.}
  \label{fig:Xlnufits}
\end{center}
\end{figure}

The fit has three yields associated with three probability density functions (pdfs), which describe the $B \to X\ell\nu$ signal decays, background from $e^{+}e^{-} \rightarrow q \bar{q}$ events, and background in which the lepton is fake or secondary. ``Secondary" here refers to the situation in which the lepton is not produced directly in the decay of the $B$ meson but rather through a secondary cascade decay of a charmed meson. Meanwhile, ``Fake" refers to the case in which a hadron is mis-reconstructed as a lepton. The $B \to X\ell\nu$ signal pdf has four sub-components, which include $B \to D^{*} \ell \nu$, $B \to D\ell\nu$, $B \to X_{u} \ell \nu$ and any remaining $B \to X_{c} \ell \nu$ decays ($B \to D^{**} \ell \nu$ and $B \to D^{(*)} n \pi \ell \nu$). The relative contributions of these four components are parametrised by three fractions ($f_{D}$, $f_{D^{*}}$ and $f_{X_{u}}$).

The last term, $(k - k_{\rm constraints})^{T} \Sigma^{-1}_{\rm constraints} (k - k_{\rm constraints})$, in Equation~\ref{eq:NLL} allows for constraints on parameters in the fit. The parameter vector $k= (N(e^{+}e^{-} \rightarrow q \bar{q}), f_{D}, f_{D^{*}}, f_{X_{u}})$ contains the subset of fit parameters, which are subject to constraints. The vector $k_{\rm constraints}$ contains the corresponding nominal values to which these parameters are constrained. The continuum yield, $N(e^{+}e^{-} \rightarrow q\bar{q})$, is constrained to its expectation based on counting off-resonance events and scaling up to account for luminosity. The constraints on the three fractions are obtained from MC expectation after all branching fraction corrections are made.

\begin{figure}[h!]
\begin{center}
 \includegraphics[width=7cm]{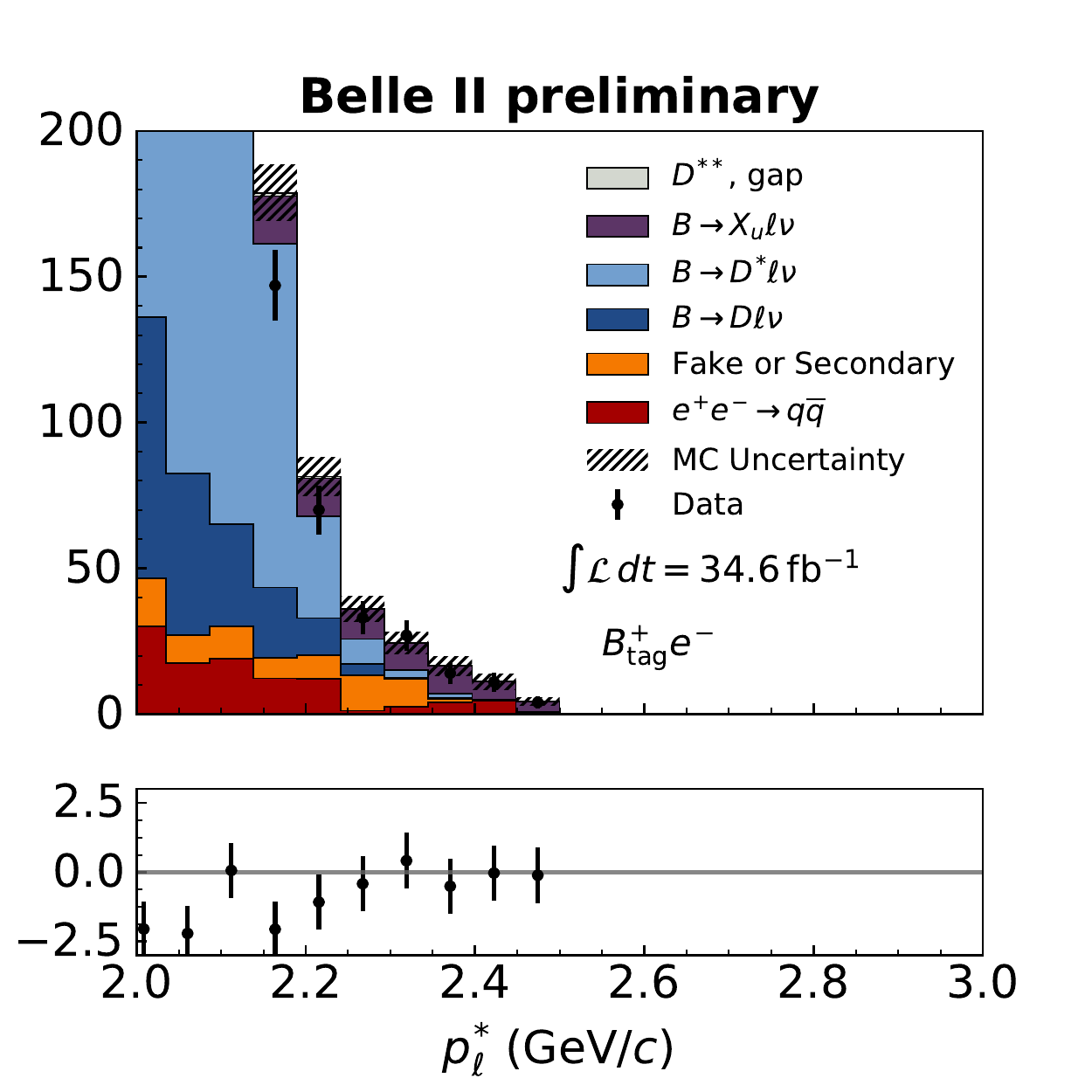} \includegraphics[width=7cm]{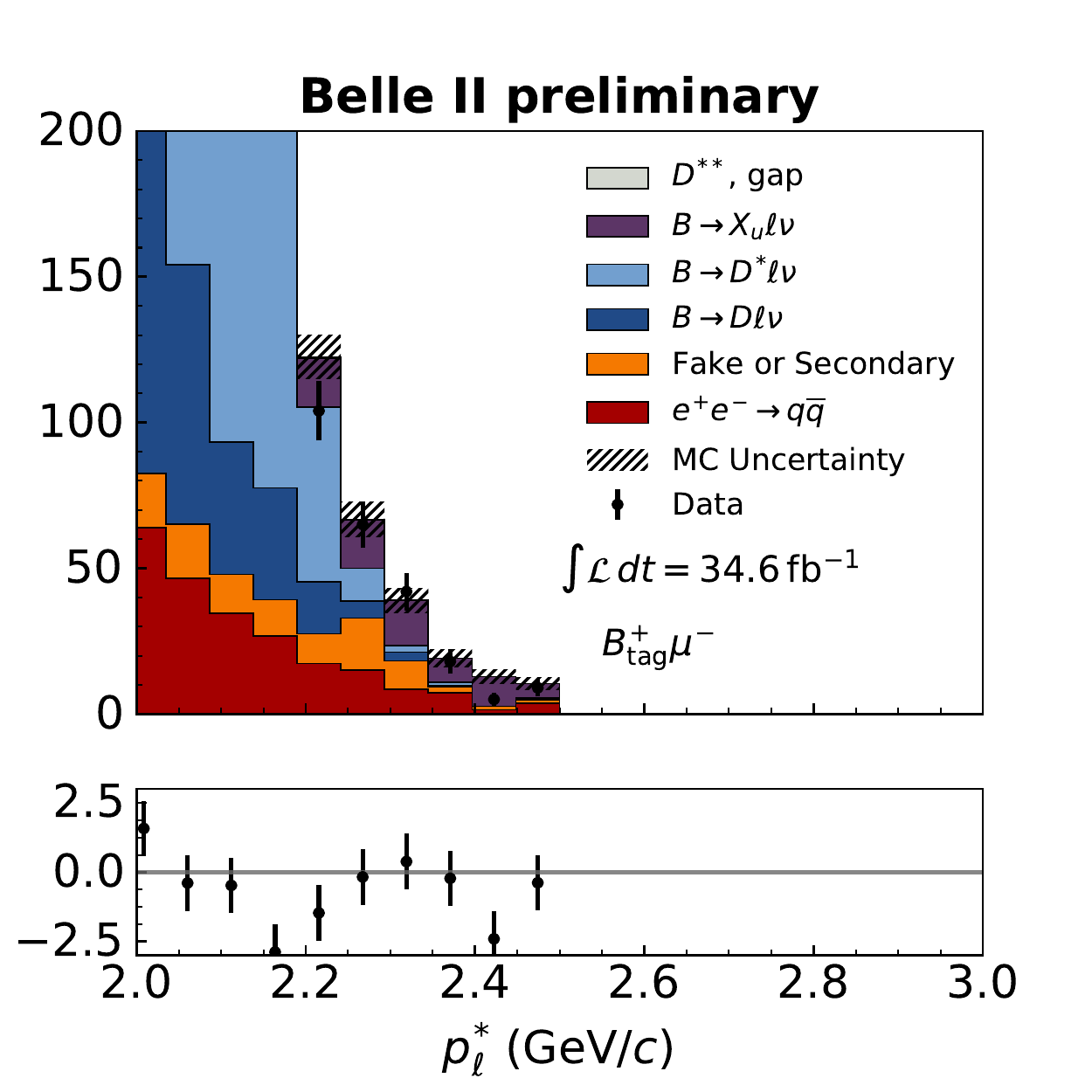} \\
	\caption{Fits to $p^{*}_{\ell}$ in data in the region $p^{*}_{\ell} > 2$~GeV/$c$. This region is enhanced in $B \rightarrow X_{u} \ell \nu$ decays relative to $B \rightarrow X_{c} \ell \nu$ decays due to the lower kinematic endpoint for $B \rightarrow X_{c} \ell \nu$ decays. }
  \label{fig:Xlnufitszoomed}
\end{center}
\end{figure}

Fit results for the channels $B^{+} e^{-}$, $B^{+} \mu^{-}$, $B^{0} e^{-}$ and $B^{0} \mu^{-}$ with a selection of $\mathcal{P} > 0.001$ are shown in Fig.~\ref{fig:Xlnufits}. A good agreement between data and the fitted models is observed across all channels. Fig.~\ref{fig:Xlnufitszoomed} shows the $B^{+} \ell^{-}$ fit channels in the region where $p^{*}_{\ell} > 2$~GeV/$c$. In this region, the contribution from $B \rightarrow X_{u} \ell \nu$ decays becomes evident due to the lower kinematic endpoint of $B \rightarrow X_{c} \ell \nu$  decays. This allows one to better constrain the albeit small contribution from $B \to X_{u} \ell \nu$ decays.

%% file: systematics.tex
% File info
%   FILE:       calibration.tex
%   FILE OWNER: William Sutcliffe
%   CONTENT:    Report on FEI calibration studies. Subsections for each separate author/
%               calibration study. 

\section{Sources of systematic uncertainty}
\label{sec:systematics}

The calibration procedure is affected by a number of sources of systematic uncertainty. These can influence the determination of the MC expected yield (normalisation uncertainties) or the shapes of pdfs in the fitting procedure (shape uncertainties). 

We first discuss the estimation of systematic uncertainties for the MC expected yield, $N^{\rm MC}_{X\ell\nu}$. The first source of systematic uncertainty considered is that arising from the knowledge of the $B \rightarrow X\ell \nu$ branching fractions. Several branching fractions of the $B \rightarrow X\ell\nu$ decay modes, including $B \to D\ell\nu$, $B \to D^{*}\ell\nu$ and $B \to X_{u} \ell \nu$, were first corrected to their latest PDG values. After having applied these corrections, the overall charged and neutral $B \rightarrow X\ell\nu$ branching fractions were scaled to match those in the PDG: $\mathcal{B}(B^{+} \rightarrow X \ell \nu ) = 10.99 \pm 0.28 $ and $\mathcal{B}(B^{0} \rightarrow X \ell \nu ) = 10.33 \pm 0.28 $. The corresponding uncertainties are treated as a source of systematic uncertainty. In addition to correcting several branching fractions, the form factors of $D \ell \nu$ and $D^{*} \ell \nu$ decays are updated to the BGL parametrisations of Ref.~\cite{Grinstein:2017nlq,Bigi:2017njr}, with the central parameter values in Ref.~\cite{Dstlnu}. The associated uncertainties on the form factor parameters of these parameterisations are propagated in the analysis using one-sigma variations in an uncorrelated eigenbasis of form factor parameters of the corresponding BGL parametrisations. The form factor uncertainties can influence $N^{\rm MC}_{X\ell\nu}$ due to the selection of $p^{*}_{\ell} > 1$~GeV/$c$.

The next sources of uncertainty relate to tracking and particle identification. Due to mismatches in the reconstruction of tracks between simulation and data, a systematic error of 0.91\% is assigned for the single signal-side track. The performance of lepton identification also differs between data and MC. Consequently, the lepton identifcation rates and $\pi \rightarrow \ell$ and $K \rightarrow \ell$ fake rates are corrected in bins of lepton momentum and polar angle using corrections derived from data samples of $J /\psi \rightarrow \ell^{+} \ell^{-}$, $D^{*+} \rightarrow (D^{0} \rightarrow  K^{-} \pi^{+}) \pi^{+}$ and $K^{0}_{S} \rightarrow \pi^{+} \pi^{-}$ decays. The systematic uncertainty associated with these corrections is determined by generating gaussian variations on these weights according to their systematic and statistical uncertainties, while assuming that the systematic uncertainties across bins are 100\% correlated. The final considered source of systematic uncertainty on $N^{\rm MC}_{X\ell\nu}$ is the statistical size of the MC sample used to estimate $N^{\rm MC}_{X\ell\nu}$.

A number of systematic effects can impact the expected $p^{*}_{\ell}$ distribution from simulation. These include the Monte Carlo statistics, the $B \to D^{(*)} \ell \nu$ form factors, lepton identification and the composition of $B \to X \ell \nu$ decays. The uncertainty associated with the composition of $B \to X \ell \nu$ is propagated into the fit through the freedom of the $B \to X \ell \nu$ pdf to change according to aforementioned sub-pdf fractions. A multivariate Gaussian constraint on these fractions is estimated, which accounts for the PDG uncertainty on several branching fraction updates and Monte Carlo statistics. Given that the contribution from $B \to D^{**} \ell \nu$ and $B \to D^{(*)} n \pi \ell \nu$ is not very well known, the overall branching fraction of these transitions is assigned a 20\% uncertainty. 

The shape impact for the remaining systematic sources of uncertainty are accounted for by using the nuisance parameters associated with each bin of a sub-pdf. For each systematic source of uncertainty, $s$, a $N_{\rm dim} \times N_{\rm dim}$ covariance matrix, $\Sigma_{s}$, is estimated, where $N_{\rm dim} = N_{\rm bins} \times N_{\rm pdfs}$. For lepton identification, $\Sigma_{\rm LID}$, is estimated by filling histograms with each independent weight variation. Meanwhile, for the $D^{(*)}$ form factors, $\Sigma_{D^{(*)} {\rm FF}}$ is estimated by combining covariance matrices associated with one-sigma eigen-variations of BGL form factor parameters. Lastly, for MC statistics, $\Sigma_{\rm MC}$ is determined using Poisson statistics and is purely diagonal. The total covariance matrix $\Sigma_{\theta} = \sum_{s} \Sigma_{s}$ is used in the nuisance parameter constraint term of Equation~\ref{eq:NLL}.

%% file: results.tex
% File info
%   FILE:       calibration.tex
%   FILE OWNER: William Sutcliffe
%   CONTENT:    Report on FEI calibration studies. Subsections for each separate author/
%               calibration study. 

\section{results}
\label{sec:results}

Final results for the calibration factors as determined from the fitted yields are shown in Fig.~\ref{fig:calfig}. The corresponding numerical results are itemised in Appendix~\ref{sec:appendix} along with the simulated and fitted yields of $B \to X\ell\nu$ decays. Calibration factors for tag-side $B^{0}$ and $B^{+}$ mesons are found to agree well for both lepton channels with the $B^{+}$ and $B^{0}$ calibration factors ranging from $0.60$-$0.63$ and $0.70$-$0.83$, respectively. For tag-side $B^{0}$ mesons, the calibration factors with a looser selection on the tag-side $B$ classifier output, $\mathcal{P}_{B^{0}_{\rm tag}}$, are generally observed to be higher. This appears to be due to the fact that a looser cut increases the contribution of certain modes in the lower purity region. The sources uncertainties for the calibration factors are shown in Table~\ref{table:sys} for the threshold of $\mathcal{P} > 0.001$. The dominant systematic uncertainty is associated with the shape freedom in the fit, which ranges from $2$ to $4\%$, depending on the channel. The next largest sources of uncertainty are those associated with $\mathcal{B}(B^{+/0} \rightarrow X \ell \nu)$ ($2.1\%$) and tracking ($0.91\%$).

The calibration factors are subsequently averaged across lepton modes as displayed in Table~\ref{table:final} and in Fig.~\ref{fig:calfig}. The averaging procedure uses a weighted average, that accounts for the relative uncertainties and correlations of the measurements. In particular, the uncertainties from tracking, $\mathcal{B}(B^{+/0} \rightarrow X \ell \nu)$, and the $D^{(*)}\ell\nu$ form factors are deemed to be 100\% correlated. 

\begin{figure}[h!]
\begin{center}
	\includegraphics[width=8cm]{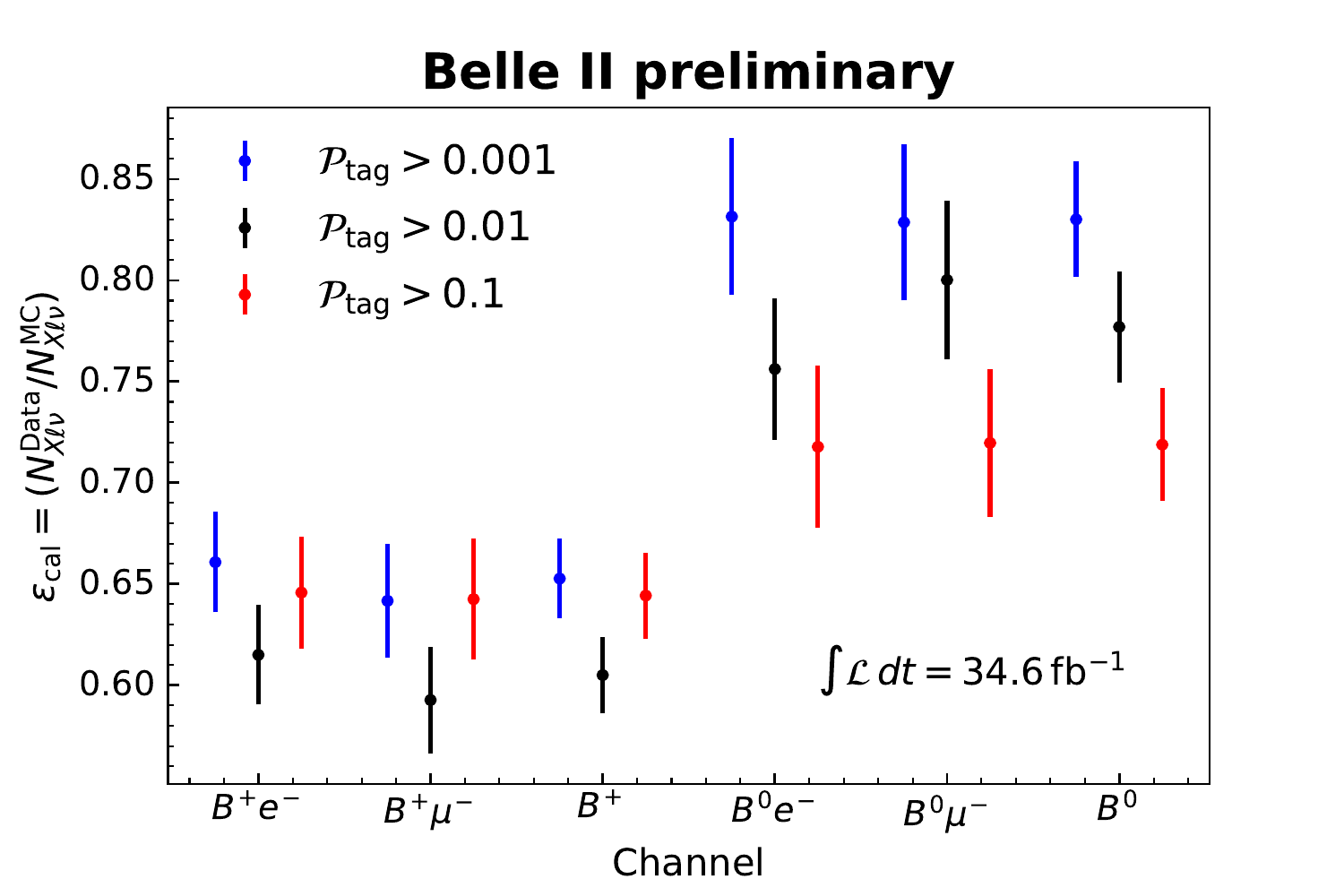} \put(-230,130){(a)}
	 \includegraphics[width=8cm]{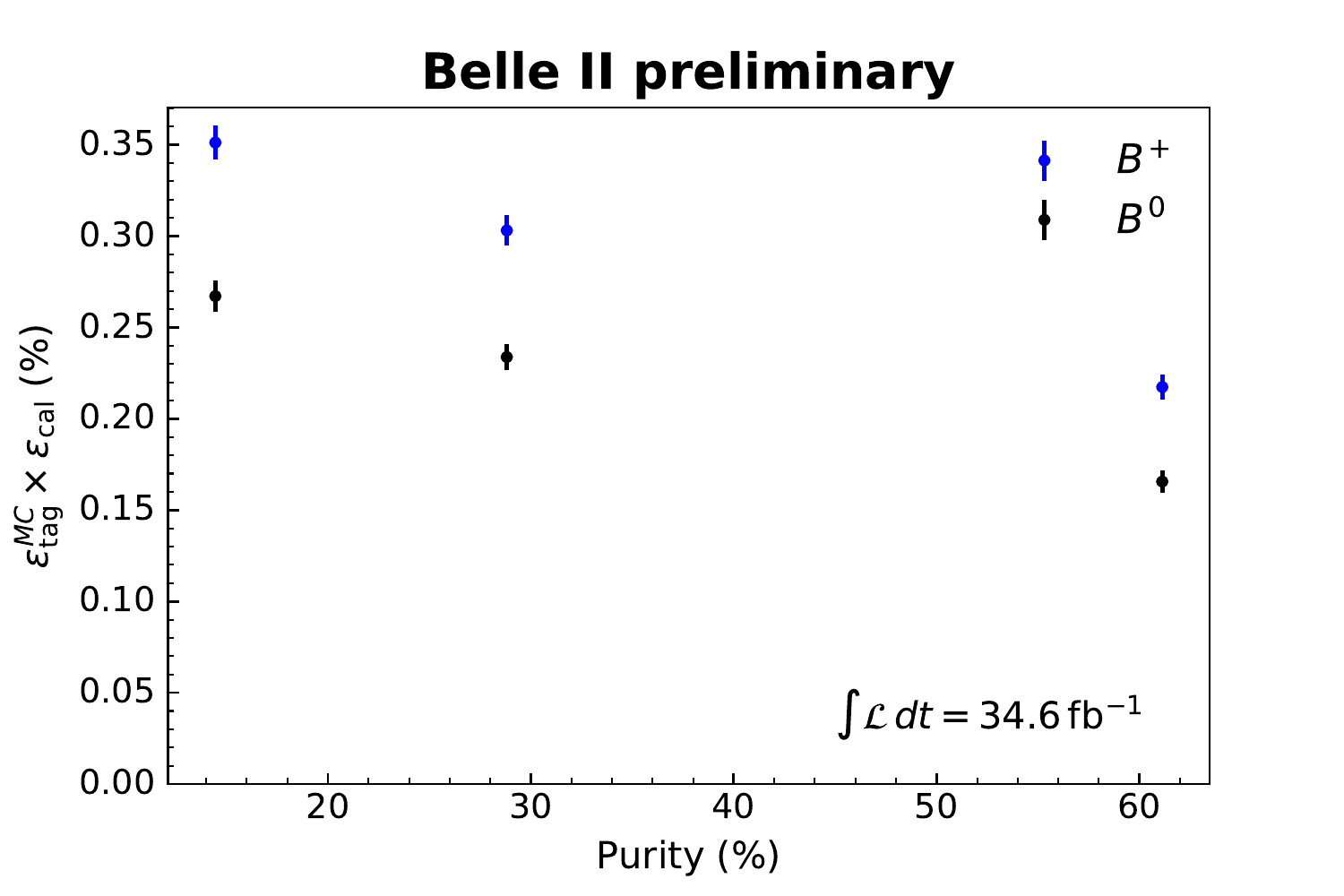} \put(-230,130){(b)}\\
\end{center}
	\caption{(a) Calibration factors for each of the different channels and different signal probability, $\mathcal{P}_{\rm tag}$, selection choices. Good agreement is seen between the muon and electron channels for the signal-side $B \rightarrow X \ell \nu$ decay. (b) $\epsilon^{\rm MC}_{\rm tag} \times \epsilon_{\rm cal}$ against purity for $\mathcal{P}_{\rm tag} > 0.001$, $0.01$ and $0.1$ for $B^{0}$ and $B^{+}$ mesons.}
  \label{fig:calfig}
\end{figure}

\begin{table}[!h]
\begin{tabular}{lccccccccc}
\hline
	Channel & MC Stat. & $\mathcal{B}(B^{0/+}\rightarrow X\ell\nu)$ & Tracking & $D\ell\nu$ FF & Lepton ID & $D^{*}\ell\nu$ FF & Fit Stat. & Fit Model\\
\hline
$B^{+}e^{-}$ & 0.39 & 2.09 & 0.91 & 0.06 & 0.76 & 0.41 & 0.93 & 2.67 \\
$B^{+}\mu^{-}$ & 0.37 & 2.1 & 0.91 & 0.06 & 2.13 & 0.38 & 0.86 & 2.93 \\
$B^{0}e^{-}$ & 0.62 & 2.1 & 0.91 & 0.07 & 0.73 & 0.43 & 1.22 & 3.72 \\
$B^{0}\mu^{-}$ & 0.6 & 2.09 & 0.91 & 0.06 & 2.13 & 0.41 & 1.19 & 3.17 \\
\hline
\end{tabular}
	\caption{Itemisation of the percentage contribution from the sources of uncertainty on the calibration factors for the selection $\mathcal{P}_{\rm tag} > 0.001$.}
  \label{table:sys}
\end{table}

\begin{table}
\begin{tabular}{lcc}
\hline
	$B^{+}$\\
\hline
$\mathcal{P}_{\rm tag}$ $>$ & $\epsilon$ & uncertainty [\%]\\
\hline
0.001 & $0.65 \pm 0.02$ & 3.0\\
0.01 & $0.61 \pm 0.02$ & 3.1\\
0.1 & $0.64 \pm 0.02$ & 3.3\\
\hline
\end{tabular}

\begin{tabular}{lcc}
\hline
	$B^{0}$\\
\hline
$\mathcal{P}_{\rm tag}$ $>$ & $\epsilon$ & uncertainty [\%]\\
\hline
0.001 & $0.83 \pm 0.03$ & 3.4\\
0.01 & $0.78 \pm 0.03$ & 3.5\\
0.1 & $0.72 \pm 0.03$ & 3.9\\
\hline
\end{tabular}

	\caption{Final calibration factors averaged over lepton type. A weighted average taking into account the uncertainties and correlated systematics is used. }
  \label{table:final}
\end{table}

The final calibration factors, $\epsilon_{\rm cal}$, in Table~\ref{table:final} can be applied in order to correct the tag-side efficiency in simulation, $\epsilon^{\rm MC}_{\rm tag}$. In Fig.~\ref{fig:calfig} the corrected tag-side efficiency from simulation, $\epsilon^{\rm MC}_{\rm tag} \times \epsilon_{\rm cal}$, is shown against purity, for the $\mathcal{P}_{\rm tag}$ thresholds of $0.001$, $0.01$ and $0.1$. Here, the tag-side efficiency, $\epsilon^{\rm MC}_{\rm tag}$, refers to ratio of the number of events containing a correctly reconstructed tag-side $B$ meson in the region $M_{bc} > 5.27$~GeV/$c^{2}$ to the total number of simulated $\Upsilon(4S) \rightarrow B\bar{B}$ events. Meanwhile the purity is the ratio of the number of events containing a correctly reconstructed tag-side $B$ meson in this region to the number of events containing a reconstructed tag-side $B$ meson.

%% file: conclusions.tex
% File info
%   FILE:       conclusions.tex
%   FILE OWNER: Peter Lewis
%   CONTENT:    Upshot of FEI performance for early phase III

\section{Conclusions}
\label{sec:conclusions}

At Belle II, hadronic tag-side reconstruction will be a critical part of the physics program, allowing a number of challenging final states with missing energy to be measured. This includes measurements of $R(D^{(*)})$ with $B \rightarrow D^{(*)} \tau \nu$ decays, measurements of the CKM matrix elements $|V_{ub}|$ and $|V_{cb}|$ using inclusive $B \rightarrow X_{c/u} \ell \nu$ transitions and searches for the rare decay $B \rightarrow K^{*} \nu \bar{\nu}$.

The Belle II experiment's tag-side reconstruction algorithm, Full Event Interpretation, relies on a hierarchical reconstruction of around 10000 $B$ meson decays with over 200 multivariate classifiers. In order to employ the algorithm in a physics analysis, it is necessary to account for differences in the performance of the algorithm between data and simulation. Here, first calibration factors were derived in order to correct for these effects by measuring a well-known signal side of $B\rightarrow X\ell\nu$ decays. Calibration factors are determined for both $B^{0}$ and $B^{+}$ mesons for a range of selections on the tag-side $B$ multivariate classifier. For a very loose selection, the calibration factors are $0.653 \pm 0.020$ and $0.830 \pm 0.029$ for tag-side $B^{+}$ and $B^{0}$ mesons, respectively. 

%% file: acknowledgements.tex
We thank the SuperKEKB group for the excellent operation of the
accelerator; the KEK cryogenics group for the efficient
operation of the solenoid; and the KEK computer group for
on-site computing support.
This work was supported by the following funding sources:
%Armenia
Science Committee of the Republic of Armenia Grant No. 18T-1C180;
%Australia
Australian Research Council and research grant Nos.
DP180102629, 
DP170102389, 
DP170102204, 
DP150103061, 
FT130100303, 
and
FT130100018; 
%Austria
Austrian Federal Ministry of Education, Science and Research, and
Austrian Science Fund No. P 31361-N36; 
%Canada
Natural Sciences and Engineering Research Council of Canada, Compute Canada and CANARIE;
%China
Chinese Academy of Sciences and research grant No. QYZDJ-SSW-SLH011,
National Natural Science Foundation of China and research grant Nos.
11521505,
11575017,
11675166,
11761141009,
11705209,
and
11975076,
LiaoNing Revitalization Talents Program under contract No. XLYC1807135,
Shanghai Municipal Science and Technology Committee under contract No. 19ZR1403000,
Shanghai Pujiang Program under Grant No. 18PJ1401000,
and the CAS Center for Excellence in Particle Physics (CCEPP);
%Czech Republic
the Ministry of Education, Youth and Sports of the Czech Republic under Contract No.~LTT17020 and 
Charles University grants SVV 260448 and GAUK 404316;
%EU
European Research Council, 7th Framework PIEF-GA-2013-622527, 
Horizon 2020 Marie Sklodowska-Curie grant agreement No. 700525 `NIOBE,' 
and
Horizon 2020 Marie Sklodowska-Curie RISE project JENNIFER2 grant agreement No. 822070 (European grants);
%France
L'Institut National de Physique Nucl\'{e}aire et de Physique des Particules (IN2P3) du CNRS (France);
%Germany
BMBF, DFG, HGF, MPG, AvH Foundation, and Deutsche Forschungsgemeinschaft (DFG) under Germany's Excellence Strategy -- EXC2121 ``Quantum Universe''' -- 390833306 (Germany);
%India
Department of Atomic Energy and Department of Science and Technology (India);
%Israel
Israel Science Foundation grant No. 2476/17
and
United States-Israel Binational Science Foundation grant No. 2016113;
%Italy
Istituto Nazionale di Fisica Nucleare and the research grants BELLE2;
%Japan
Japan Society for the Promotion of Science,  Grant-in-Aid for Scientific Research grant Nos.
16H03968, 
16H03993, 
16H06492,
16K05323, 
17H01133, 
17H05405, 
18K03621, 
18H03710, 
18H05226,
19H00682, % Niigata
26220706,
and
26400255,
the National Institute of Informatics, and Science Information NETwork 5 (SINET5), 
and
the Ministry of Education, Culture, Sports, Science, and Technology (MEXT) of Japan;  
%Korea
National Research Foundation (NRF) of Korea Grant Nos.
2016R1\-D1A1B\-01010135,
2016R1\-D1A1B\-02012900,
2018R1\-A2B\-3003643,
2018R1\-A6A1A\-06024970,
2018R1\-D1A1B\-07047294,
2019K1\-A3A7A\-09033840,
and
2019R1\-I1A3A\-01058933,
Radiation Science Research Institute,
Foreign Large-size Research Facility Application Supporting project,
the Global Science Experimental Data Hub Center of the Korea Institute of Science and Technology Information
and
KREONET/GLORIAD;
%Malaysia
Universiti Malaya RU grant, Akademi Sains Malaysia and Ministry of Education Malaysia;
%Mexico
% CINVESTAV-IPN, UNAM, UAS, BUAP and CONACYT are funded under
Frontiers of Science Program contracts
FOINS-296,
CB-221329,
CB-236394,
CB-254409,
and
CB-180023, and SEP-CINVESTAV research grant 237 (Mexico);
%Poland
the Polish Ministry of Science and Higher Education and the National Science Center;
%Russia
the Ministry of Science and Higher Education of the Russian Federation,
Agreement 14.W03.31.0026;
%Saudi Arabia
University of Tabuk research grants
S-1440-0321, S-0256-1438, and S-0280-1439 (Saudi Arabia);
%Slovenia
Slovenian Research Agency and research grant Nos.
J1-9124
and
P1-0135; 
%Spain
Agencia Estatal de Investigacion, Spain grant Nos.
FPA2014-55613-P
and
FPA2017-84445-P,
and
CIDEGENT/2018/020 of Generalitat Valenciana;
%Taiwan
Ministry of Science and Technology and research grant Nos.
MOST106-2112-M-002-005-MY3
and
MOST107-2119-M-002-035-MY3, 
and the Ministry of Education (Taiwan);
%Thailand
Thailand Center of Excellence in Physics;
%Turkey
TUBITAK ULAKBIM (Turkey);
%Ukraine
Ministry of Education and Science of Ukraine;
%USA
the US National Science Foundation and research grant Nos.
PHY-1807007 % Luther
and
PHY-1913789, % Indiana CEEM
and the US Department of Energy and research grant Nos.
DE-AC06-76RLO1830, % PNNL
DE-SC0007983, % Wayne State
DE-SC0009824, % Florida
DE-SC0009973, % VPI
DE-SC0010073, % South Carolina
DE-SC0010118, % Carnegie Mellon
DE-SC0010504, % Hawaii
DE-SC0011784, % Cincinnati
DE-SC0012704; % BNL
%last group
and
%Vietnam
the National Foundation for Science and Technology Development (NAFOSTED) 
of Vietnam under contract No 103.99-2018.45.

%% file: appendix.tex
\appendix
\section{}
\label{sec:appendix}

A summary of all fitted yields, $N^{\rm Data}_{X\ell\nu}$, MC expected yields, $N^{\rm MC}_{X\ell\nu}$ and the corresponding calibration factors are provided in Table~\ref{table:calresults}. 

\begin{table}[!h]
\begin{tabular}{lccc}
\hline 
Sig. Prob. $>0.001 $\\
\hline
Channel & $N^{\rm MC}_{X\ell\nu}$ & $N^{\rm Data}_{X\ell\nu}$ & $\epsilon$\\
\hline
	$B^{+}e^{-}$ & $(4.46 \pm 0.11) \times 10^{4}$ & $ (2.94 \pm 0.08) \times 10^{4}$ & $0.66 \pm 0.02$\\
	$B^{+}\mu^{-}$ & $(4.78 \pm 0.11) \times 10^{4} $ & $ (3.10 \pm 0.10) \times 10^{4} $ & $0.65 \pm 0.03$\\
	$B^{0}e^{-}$ & $(1.75 \pm 0.04) \times 10^{4}$ & $ (1.46 \pm 0.07) \times 10^{4}$ & $0.83 \pm 0.04$\\
$B^{0}\mu^{-}$ & $( 1.85 \pm 0.06) \times 10^{4}$ & $ (1.54 \pm 0.05) \times 10^{4} $ & $0.83 \pm 0.04$\\
\hline
\end{tabular}
\begin{tabular}{lccc}
\hline 
Sig. Prob. $>0.01 $\\
\hline
Channel & $N^{\rm MC}_{X\ell\nu}$ & $N^{\rm Data}_{X\ell\nu}$ & $\epsilon$\\
\hline
	$B^{+}e^{-}$ & $(2.65 \pm 0.07 ) \times 10^{4}$ & $ (1.63  \pm 0.05) \times 10^{4} $ & $0.62 \pm 0.02$\\
	$B^{+}\mu^{-}$ & $(2.88 \pm 0.09) \times 10^{4} $ & $(1.71 \pm 0.05 )\times 10^{4} $ & $0.59 \pm 0.03$\\
	$B^{0}e^{-}$ & $(1.11 \pm 0.03) \times 10^{4} $ & $(0.84 \pm 0.04 )\times 10^{4}$ & $0.76 \pm 0.04$\\
	$B^{0}\mu^{-}$ & $(1.18 \pm 0.04)\times 10^{4} $ & $(0.94 \pm 0.03 )\times 10^{4}$ & $0.80 \pm 0.04$\\
\hline
\end{tabular}
\begin{tabular}{lccc}
\hline 
Sig. Prob. $>0.1 $\\
\hline
Channel & $N^{\rm MC}_{X\ell\nu}$ & $N^{\rm Data}_{X\ell\nu}$ & $\epsilon$\\
\hline
	$B^{+}e^{-}$ & $ (1.10  \pm 0.03 ) \times 10^{4} $ & $(0.71  \pm 0.03 ) \times 10^{4} $ & $0.65 \pm 0.03$\\
	$B^{+}\mu^{-}$ & $(1.21 \pm 0.04) \times 10^{4} $ & $(0.78  \pm 0.04 ) \times 10^{4} $ & $0.64 \pm 0.03$\\
$B^{0}e^{-}$ & $(0.60  \pm 0.02 ) \times 10^{4} $ & $ (0.43  \pm 0.02 ) \times 10^{4} $ & $0.72 \pm 0.04$\\
$B^{0}\mu^{-}$ & $(0.64  \pm 0.02 ) \times 10^{4} $ & $(0.46  \pm 0.02 ) \times 10^{4} $ & $0.72 \pm 0.04$\\
\hline
\end{tabular}
	\caption{Results for $N_{X \ell\nu}$ as determined from the fits to data and simulation together with total uncertainties. The corresponding calibration factors computed from the ratio of these yields are also shown for each channel.  }
  \label{table:calresults}
\end{table}